\documentclass[a4paper,11pt]{article}

\pdfoutput=1 

\usepackage{jcappub} 
\usepackage{amsmath}
\usepackage{amsfonts}
\usepackage{amssymb}
\usepackage{graphicx}
\def\figurename{Fig.}

\title{\boldmath More accurate slow-roll approximations for inflation in scalar-tensor theories}

\author[a]{Ekaterina~O.~Pozdeeva,}
\author[b]{Maria~A.~Skugoreva,}
\author[b]{Alexey~V.~Toporensky,}
\author[a,1]{Sergey~Yu.~Vernov\note{The corresponding author.}}

\affiliation[a]{Skobeltsyn Institute of Nuclear Physics, Lomonosov Moscow State University,\\
Leninskiye Gory~1, Moscow 119991, Russia}
\affiliation[b]{Sternberg Astronomical Institute, Lomonosov Moscow State University, Universitetsky Prospect~13, Moscow 119991, Russia}

\emailAdd{pozdeeva@www-hep.sinp.msu.ru}
\emailAdd{masha-sk@mail.ru}
\emailAdd{atopor@rambler.ru}
\emailAdd{svernov@theory.sinp.msu.ru}


\arxivnumber{2502.13008}

\abstract{ We propose new versions of the slow-roll approximation for inflationary models with nonminimally coupled scalar fields. We derive more precise expressions for the standard slow-roll parameters as functions of the scalar field. To verify the accuracy of the proposed approximations, we consider inflationary models with the induced gravity term and the fourth-order monomial potential. For specific values of the model parameters, this model is the well-known Higgs-driven inflationary model. We investigate the inflationary dynamics in the Jordan frame and come to the conclusion that the proposed versions of the slow-roll approximation are not only more accurate at the end of inflation, but also give essentially more precise estimations for the tensor-to-scalar ratio~$r$ and the amplitude of scalar perturbations~$A_s$.
}

\begin{document}
\maketitle
\flushbottom

\section{Introduction}
~~~~A simple explanation of the fact that the Universe is spatially flat, approximately isotropic and homogeneous at cosmological distances based on the existence of a stage of accelerated expansion in the very early Universe, so-called cosmological inflation. Models of inflation yield accurate quantitative predictions of parameters of relic cosmological perturbations. The main inflationary parameters:
amplitude of scalar perturbations $A_s$, their spectral index $n_s$, and the  tensor-to-scalar ratio of the density perturbations $r$ are constrained by the combined analysis of Planck~\cite{Planck:2018jri}, BICEP/Keck~\cite{BICEP:2021xfz} and other observations as follows~\cite{Galloni:2022mok}:
\begin{equation}
\label{Inflparamobserv}
A_s=(2.10\pm 0.03)\times 10^{-9},\qquad n_s=0.9654\pm 0.0040  \qquad {\rm and} \qquad  r < 0.028 \,.
\end{equation}

    The observational data constraints possible inflationary scenarios. In particular, the inflationary model with a minimally coupled scalar field and the fourth-order monomial potential~\cite{Linde:1981mu,Linde:1983gd}
that predicts rather large values of $r$ is ruled out, meanwhile the model with the induced gravity term and the same potential~\cite{Spokoiny:1984bd} is in good agreement with the restrictions (\ref{Inflparamobserv}) on the inflationary parameters. The induced gravity term naturally appears when quantum corrections are taken into account, since generic quantum corrections to the action of the scalar field minimally coupled to gravity do include this term~\cite{Chernikov:1968zm,Callan:1970ze,Tagirov:1972vv}.
Inflationary models with the Ricci scalar multiplied by a function of the scalar field are being studied intensively~\cite{Spokoiny:1984bd,Accetta:1985du,Lucchin:1985ip,
Futamase:1987ua,Makino:1991sg,Barvinsky:1994hx,Kaiser:1994vs,Cervantes-Cota:1995ehs,Barrow:1995fj,Libanov:1998wg,Faraoni:2000wk,Bezrukov:2007ep,Barvinsky:2008ia,Garcia-Bellido:2008ycs,DeSimone:2008ei,
Barvinsky:2009ii,Bezrukov:2010jz,Atkins:2010yg,Kallosh:2013pby,Bezrukov:2013fka,Bezrukov:2014ipa,Ren:2014sya,Kallosh:2014laa,Elizalde:2014xva,Pieroni:2015cma,Elizalde:2015nya,Jarv:2016sow,Kuusk:2016rso,Pozdeeva:2016hrz,
Burns:2016ric,Karam:2017zno,Odintsov:2018ggm,
Kamenshchik:2019rot,Mishra:2019ymr,Akin:2020mcr,Jarv:2021qpp,Karciauskas:2022jzd,
Diaz:2023tma,Chataignier:2023ago,Kamenshchik:2024kay}.

    The standard way to analyze an inflationary model with a nonminimally coupled scalar field starts from simplification of evolution equations by removing of terms corresponding to the nonminimal coupling. This procedure includes the conformal transformation of the metric and construction of the corresponding model in the Einstein frame.
After this, the standard formulae for the slow-roll single-field inflation with a minimally coupled scalar field are used~\cite{Liddle:1994dx,Lidsey:1995np}.

    This method is suitable to investigate $F(R)$ inflationary models~\cite{Starobinsky:1980te,Mijic:1986iv,Maeda:1987xf,Saidov:2010wx,Miranda:2017juz,Odintsov:2020thl,Ivanov:2021chn,Odintsov:2022rok,Pozdeeva:2022lcj} as well, but it cannot be used for construction of inflationary models that include more complicated geometrical structures, for example the Gauss-Bonnet term~\cite{Guo:2010jr,Koh:2014bka,vandeBruck:2015gjd,Kanti:2015pda,Koh:2016abf,vandeBruck:2017voa,Odintsov:2018zhw,Yi:2018gse,Granda:2019wpe,
Pozdeeva:2020apf,Pozdeeva:2020shl,Pozdeeva:2021nmz,Pozdeeva:2021iwc,Oikonomou:2021kql,Kawai:2021edk,Ketov:2022lhx,Ketov:2022zhp,Koh:2023zgn,Pozdeeva:2024ihc,Yogesh:2024zwi,Pozdeeva:2024kzb}.
To calculate the inflationary parameters one uses the slow-roll approximation that essentially simplifies the search and analysis of inflationary models.
The investigation of the slow-roll dynamics in the Jordan frame only can be considered as the first step for the construction of slow-roll approximations for more complicated cosmological models.

    For the General Relativity (GR) models with minimally coupled scalar fields, the slow-roll approximation is equivalent to the assumption that all slow-roll parameters are small in comparison with unity. These parameters are defined as functions of the Hubble parameter and the scalar field. In the slow-roll approximation, both the Hubble parameter and slow-roll parameters are functions of the scalar field only. By this reason, the slow-roll approximation essentially simplifies the search for model parameters suitable for inflation.  So, it is important to get such expressions of the Hubble parameter and slow-roll parameters via the scalar field that they are sufficiently close to numerical solutions of the evolution equations without any approximations.

    In this paper, we construct and compare slow-roll approximations for models with one nonminimally coupled scalar field. There are two known versions of the Jordan frame slow-roll approximation for such inflationary models. The first variant of the approximation can be considered directly in the Jordan frame, neglecting all terms containing slow-roll parameters (so that, assuming that all ones are  negligibly small). The second one uses the transition to the Einstein frame (see below). It was already remarked that these two approaches lead to two different approximations~\cite{Akin:2020mcr}. By direct comparison with solutions of numerically integrated exact equations of motion, it has been shown that the second way gives an essentially better approximation~\cite{Jarv:2021qpp}. In our paper, we show that the known expressions of the Hubble parameter and slow-roll parameters via the scalar field can be improved, proposing new slow-roll approximations. These approximations are more accurate than the known ones not only at the end of inflation, but also at the beginning, that allows one to get the value of parameter $r$ with essentially more accuracy.

    The most popular single-field inflationary model with nonminimal coupling includes the induced gravity term and fourth-order monomial potential. This model was proposed in 1984 in Ref.~\cite{Spokoiny:1984bd} and developed in Refs.~\cite{Makino:1991sg,Barvinsky:1994hx}. The additional assumption that the inflaton is the Standard model Higgs boson has been proposed in Ref.~\cite{Bezrukov:2007ep} and gives a start to actively investigations of this model~\cite{Barvinsky:2008ia,Garcia-Bellido:2008ycs,DeSimone:2008ei,Barvinsky:2009ii,Bezrukov:2010jz,Atkins:2010yg,Kallosh:2013pby,Bezrukov:2013fka,Bezrukov:2014ipa,Mishra:2019ymr}.
 This model has been used to test the accuracy of the known slow-roll approximations in the Jordan frame~\cite{Jarv:2021qpp}. In the present paper, we use this model with different values of parameters to compare the proposed approximations with the known ones.

    The paper is organized as follows. In Section~2, we remind the evolution equations. The slow-roll parameters are defined in Section~3. The known versions of the slow-roll approximation are presented in Section~4. In Section~5, we propose new variants of the approximated equations. All these approximations are compared with the results of numerical integration without any approximations in Section~6. The obtained results are summarized in Section~7. Explicit expressions for the slow-roll parameters $\varepsilon_2$ and $\zeta_2$ as functions of the scalar field $\phi$ are presented in Appendix~A.
 In Appendix~B, the alternative way to get the approximation II is presented.

\section{Model with the nonminimal coupling}
~~~~The action of a generic model with a nonminimally coupled scalar field $\phi$,
\begin{equation}
S=\frac12\int d^4x\sqrt{-g}\left(F(\phi)R-g^{\mu\nu}\partial_\mu\phi\partial_\nu\phi-2V(\phi)\right),
\label{action1}
\end{equation}
includes the coupling function $F(\phi)>0$ and the potential $V(\phi)$.

    In the spatially flat Friedmann-Lema\^{i}tre-Robert\-son-Walker (FLRW) metric with
\begin{equation}
\label{FLRW}
ds^2={}-dt^2+a^2(t) \, d\mathbf{x}^2 \,,
\end{equation}
the system of the field equations is
\begin{equation}
\label{equ00}
3H^2F=\frac{1}{2}{\dot\phi}^2+V-3F_{,\phi}\dot\phi H,
\end{equation}
\begin{equation}
\label{dotH}
2\dot H F={}-{\dot\phi}^2+F_{,\phi}\dot\phi H-F_{,\phi\phi}{\dot\phi}^2-F_{,\phi}\ddot\phi,
\end{equation}
\begin{equation}
\label{KG}
\ddot\phi+3H\dot\phi+ V_{,\phi}-3F_{,\phi}\left(\dot H+2H^2\right)=0,
\end{equation}
where dots denote the time derivatives and commas denote  derivatives with respect to the scalar field, $A_{,\phi}=\tfrac{d A(\phi)}{d \phi}$ for any function $A$.

    It is suitable to present Eqs.~(\ref{dotH}) and (\ref{KG}) as the following dynamical system:
\begin{equation}
\label{DynSYS}
\begin{split}
\dot\phi=&\psi,\\
\dot\psi=&\frac{1}{E}\left\{{}-3F_{,\phi}\left[F_{,\phi\phi}+1\right]\psi^2+3\left[F_{,\phi}^2-2F\right]H\psi-2F\left[V_{,\phi}-6F_{,\phi}H^2\right]\right\},\\
\dot H=&\frac{1}{E}\left\{{}-\left[F_{,\phi\phi}+1\right]\psi^2+4F_{,\phi}H\psi+F_{,\phi}\left[V_{,\phi}-6F_{,\phi}H^2\right]\right\},
\end{split}
\end{equation}
where
\begin{equation}
E=3F_{,\phi}^2+2F.
\end{equation}

    We consider the e-folding number $N=\ln(a/a_{e})$, where $a_{e}$ is a constant,  as a measure of time during inflation. Using the relation $\frac{d}{dt}=H\,\frac{d}{dN}$ and introducing the function $\chi(N)=\frac{d\phi}{dN}$,  we can write system~(\ref{DynSYS}) as follows:
\begin{equation}
\label{DynSYSN}
\begin{split}
\frac{d\phi}{dN}=&\chi,\\
\frac{d\chi}{dN}=&\frac{1}{E}\left\{{}-3F_{,\phi}\left[F_{,\phi\phi}+1\right]\chi^2+3\left[F_{,\phi}^2-2F\right]\chi-2\frac{FV_{,\phi}}{H^2}+12FF_{,\phi}\right\}-\frac{\chi}{2H^2}\frac{dH^2}{dN},\\
\frac{dH^2}{dN}=&\frac{2}{E}\left\{{}-\left[F_{,\phi\phi}+1\right]H^2\chi^2+4F_{,\phi}H^2\chi+F_{,\phi}\left[V_{,\phi}-6F_{,\phi}H^2\right]\right\}.
\end{split}
\end{equation}

    Equation (\ref{equ00}) can be presented in the following form
\begin{equation}
\label{H2phichi}
H^2=\frac{2V}{6F+6F_{,\phi}\chi-\chi^2}.
\end{equation}

    We always assume positivity of the potential $V$ during inflation, so the following condition should be satisfied:
\begin{equation}
\label{condH2}
6F+6F_{,\phi}\chi-\chi^2>0.
\end{equation}

    Substituting Eq.~(\ref{H2phichi}) into system (\ref{DynSYSN}), we  eliminate $H^2$ from the right-hand side and get the following second-order system:
\begin{equation}
\label{PhiSYSN}
\begin{split}
\frac{d\phi}{dN}=&{ } \chi,\\
\frac{d\chi}{dN}=&{ } \frac{1}{2EV}\left\{\left[2V\left(F_{,\phi\phi}+1\right)+F_{,\phi}V_{,\phi}\right]\chi^3+2\left[FV_{,\phi}-F_{,\phi}V\left(3F_{,\phi\phi}+7\right)-3F_{,\phi}^2V_{,\phi}\right]\chi^2\right.\\
+&\left. { } 6\left[3F_{,\phi}^2V-3FF_{,\phi}V_{,\phi}-2FV\right]\chi+12F\left(2F_{,\phi}V-FV_{,\phi}\right)\right\}.\\
\end{split}
\end{equation}
A notable feature of this system is that the potential appears as the first derivative of its logarithm only. System (\ref{PhiSYSN}) is suitable to get numerical solutions of the evolution equations without any approximations. Evolution of the Hubble parameter can be described by the following equation:
\begin{equation}
\label{DH2DN}
\begin{split}
\frac{dH^2}{dN}=&{}-\frac{2}{\left(6F+6F_{,\phi}\chi-\chi^2\right)E}\times\\
&\left\{\left[2V\left(F_{,\phi\phi}+1\right)+F_{,\phi}V_{,\phi}\right]\chi^2-2F_{,\phi}\left[3F_{,\phi}V_{,\phi}+4V\right]\chi+6F_{,\phi}\left[2F_{,\phi}V-FV_{,\phi}\right]\right\}.
\end{split}
\end{equation}

    Another way to get a dynamical system from Eqs.~(\ref{equ00})--(\ref{KG}) is to rewrite them in the following form~\cite{Kamenshchik:2024kay,Skugoreva:2014gka,Pozdeeva:2016cja}:
\begin{equation}
\label{equY}
3M_{\mathrm{Pl}}^2Y^2=\frac{A}{2}{\dot\phi}^2+V_{\mathrm{eff}},
\end{equation}
\begin{equation}
\label{Fr21Qm}
\dot Y={}-\frac{A\sqrt{F}}{2M_{\mathrm{Pl}}^3}\,{\dot\phi}^2,
\end{equation}
where
\begin{equation}
\label{Veff}
V_{\mathrm{eff}}(\phi)=\frac{M_{\mathrm{Pl}}^4}{F^2}\,V\,,
\qquad
A(\phi)=\frac{M_{\mathrm{Pl}}^4}{F^2}\left(1+\frac{3F_{,\phi}^2}{2F}\right),
\end{equation}
\begin{equation}
\label{Ydef}
Y=\frac{M_{\mathrm{Pl}}}{\sqrt{F}}\left(H+\frac{F_{,\phi}\,\dot \phi}{2F}\right).
\end{equation}

    The field equation that is a consequence of Eqs.~(\ref{equY}) and (\ref{Fr21Qm})  takes the following form:
\begin{equation}
\label{KGequVeff}
\ddot\phi={}-3\sqrt{\frac{F}{M_{\mathrm{Pl}}^2}}Y\dot{\phi}-\frac{A_{,\phi}}{2A}{\dot{\phi}}^2-\frac{{V_{\mathrm{eff}}}_{,\phi}}{A}.
\end{equation}

Note that any nonminimally coupled model with $F(\phi)>0$ can be transformed to the corresponding Einstein frame model through a conformal transformation of the metric. In the Einstein frame, the effective potential $V_{\mathrm{eff}}$ is equal to the potential~$V$ and $Y(\phi)$ is the Hubble parameter. The condition ${V_{\mathrm{eff}}}_{,\phi}(\phi_{dS})=0$ corresponds to a de Sitter solution with a constant $\phi=\phi_{dS}$. Such solutions are stable for ${V_{\mathrm{eff}}}_{,\phi\phi}(\phi_{dS})>0$ and unstable for ${V_{\mathrm{eff}}}_{,\phi\phi}(\phi_{dS})<0$, provided the condition $F(\phi_{dS})>0$ is satisfied~\cite{Skugoreva:2014gka}.

\section{Slow-roll parameters}
~~~~In the models with one nonminimally coupled scalar field, there are two sets of the slow-roll parameters~\cite{vandeBruck:2015gjd}:
\begin{equation}
\label{eps}
\varepsilon_1={}-\frac{\dot{H}}{H^2}=\frac{d\ln(H^{-1})}{dN}={}-\frac{1}{2}\,\frac{d\ln(H^2)}{dN}\,, \qquad
\varepsilon_n=\frac{\dot{\varepsilon}_{n-1}}{H{\varepsilon}_{n-1}}=\frac{d\ln(\varepsilon_{n-1})}{dN}\,,
\end{equation}
\begin{equation}
\label{zeta}
\zeta_1=\frac{\dot{F}}{HF}=\frac{d\ln(F)}{dN},\qquad
\zeta_n=\frac{\dot{\zeta}_{n-1}}{H{\zeta}_{n-1}}=\frac{d\ln(\zeta_{n-1})}{dN}\,.
\end{equation}

    In particular,
\begin{equation}
\zeta_2=\frac{\ddot{F}}{H\dot{F}}-\zeta_1+\varepsilon_1.
\end{equation}

    We rewrite Eq.~(\ref{Ydef}) as
\begin{equation}
\label{Yzeta}
Y=\frac{M_{\mathrm{Pl}}H}{\sqrt{F\,}}\left(1+\frac{1}{2}\zeta_1\right)\,,
\end{equation}
and get Eq.~(\ref{equY}) in the following form
\begin{equation}
\label{equzeta1}
3M_{\mathrm{Pl}}^4H^2\left(1+\frac{1}{2}\zeta_1\right)^2=\frac{AF}{2}{\dot\phi}^2+FV_{\mathrm{eff}}.
\end{equation}

Equations (\ref{equ00}) and (\ref{dotH}) can be rewritten in terms of the slow-roll parameters:
\begin{equation}
\label{equ00eps1zeta1zeta2}
H^2\left(1+\zeta_1-\frac{F}{6F_{,\phi}^2}\,\zeta_1^2\right)=\frac{V}{3F},
\end{equation}
\begin{equation}
\label{dotHslrparam}
2\varepsilon_1=\frac{F}{F_{,\phi}^2}\,\zeta_1^2-\zeta_1(1-\zeta_1-\zeta_2+\varepsilon_1).
\end{equation}

    Obviously, the right-hand side of equation (\ref{equzeta1}) cannot be equal to zero if the functions $V(\phi)$ and $F(\phi)$ are positive. Therefore, if $V(\phi)>0$ and $F(\phi)>0$ for all $\phi$ in some interval $\phi_1<\phi\leqslant\phi_0$ and  $\zeta_1(\phi_0)>-2$, then  $\zeta_1(\phi)>-2$ for all $\phi$ in this interval.

    Using slow-roll equations, we get the expressions $H^2(\phi)$ and $\chi(\phi)$. These expressions allow us to obtain the slow-roll parameters as functions of $\phi$ and $N(\phi)$. For example,
\begin{equation}
\label{eps1zeta1phi}
\zeta_1=\frac{d\ln(F(\phi))}{d\phi}\,\chi(\phi),\qquad
\varepsilon_1={}-\frac{1}{2}\,\frac{d\ln(H^2(\phi))}{d\phi}\,\chi(\phi)={}-\frac{\zeta_1 F}{2F_{,\phi}}\,\frac{d\ln(H^2(\phi))}{d\phi},
\end{equation}
\begin{equation}
\label{eps2zeta2phi}
\zeta_2(\phi)=\frac{\chi(\phi)}{\zeta_1}\,\frac{d\zeta_1}{d\phi}\,, \qquad
\varepsilon_2(\phi)=\frac{d(\ln\varepsilon_1)}{d\phi}\,\chi(\phi).
\end{equation}

    The goal of this paper is to construct new slow-roll equations and compare them with the known ones. At the beginning of inflation, in particular, at the moment when the inflationary parameters are calculated, all slow-roll parameters are small. So, we can use the standard formulae to connect inflationary and slow-roll parameters. In the first order of the slow-roll parameters, the  tensor-to-scalar ratio $r$, the amplitude of scalar perturbations $A_s$, and their spectral index $n_s$ can be presented as follows~\cite{vandeBruck:2015gjd}:

\begin{eqnarray}
r&\approx&8\left|2\varepsilon_1+\zeta_1\right|=8\left|\frac{d(\ln(F/H^2))}{d\phi}\chi\right|\,, \\
  n_s&\approx&1-2\varepsilon_1-\zeta_1-\frac{2\varepsilon_1\varepsilon_2+\zeta_1\zeta_2}{2\varepsilon_1+\zeta_1}=1+\frac{r}{8}-\frac{d\ln(r)}{dN}\,,  \\
  A_s&\approx& \frac{2H^2}{\pi^2 F\, r}\,.\label{As}
\end{eqnarray}

It follows from Eq.~(\ref{dotHslrparam}) that
\begin{equation}\label{rslr2}
    r\approx 8\left|\frac{F}{F_{,\phi}^2}\,\zeta_1^2-\zeta_1(\varepsilon_1-\zeta_1-\zeta_2)\right|\,.
\end{equation}

\section{The known slow-roll approximations}

\subsection{The simplest approximation}
~~~~In the simplest approximation, one assumes that all slow-roll parameters are negligibly small and get the following  system of approximate equations~\cite{Akin:2020mcr},
\begin{equation}
\label{00apprS}
H^2\approx\frac{V}{3F},
\end{equation}
\begin{equation}
\label{GBapprS}
3H\dot\phi+ V_{,\phi}-6F_{,\phi}H^2\approx 0.
\end{equation}

    The expressions of $\chi(\phi)$ and $\zeta_1(\phi)$ are as follows:
\begin{equation}
\label{chiapprS}
\chi(\phi)=\frac{\dot\phi H}{H^2}\approx 2F_{,\phi}-\frac{FV_{,\phi}}{V},
\end{equation}
\begin{equation}
\label{zeta1apprS}
\zeta_1(\phi)=\frac{F_{,\phi}}{F}\chi(\phi)\approx \frac{2F_{,\phi}^2}{F} -\frac{F_{,\phi}V_{,\phi}}{V}.
\end{equation}

    Using Eq.~(\ref{eps1zeta1phi}), we come to
the expression
\begin{equation}
\label{eps1apprS1}
\varepsilon_1(\phi)\approx \frac{FV_{,\phi}-2F_{,\phi}V}{2V}\left(\frac{V_{,\phi}}{V}-\frac{F_{,\phi}}{F}\right).
\end{equation}

    The explicit form of the functions $\varepsilon_2(\phi)$ and $\zeta_2(\phi)$ are given in Appendix A.

\subsection{The known more accurate slow-roll approximation}
~~~~A more accurate approximation has been proposed in~Ref.~\cite{Kaiser:1994vs} (see also,~\cite{Akin:2020mcr,Jarv:2021qpp}).
It is derived using the transition between Jordan and Einstein frames. In the present paper, we do not use the notion of the Einstein frame, however, in this subsection we briefly show the derivation of the approximation considered.

    The known conformal transformation $ g_{\mu \nu}= \Omega^2 \tilde g_{\mu \nu}$ with $\Omega^2=
\frac{M_{\mathrm{Pl}}^2}{F(\phi)}$
after appropriate redefinition of the scalar field turns the initial action (\ref{action1}) to the GR action with a minimally
coupled scalar field $\varphi$. Writing the standard slow-roll approximation in the Einstein frame:

\begin{equation}
\label{H2E}
\tilde H^2=\frac{V(\varphi)}{3 M_{\mathrm{Pl}}^2},
\end{equation}

\begin{equation}
\frac{d\varphi}{d \tilde t}={ }-\frac{1}{3 \tilde H}\frac{dV(\varphi)}{d\varphi}\,,
\end{equation}
where variables with tildes  are Einstein frame variables,  and going back to the initial Jordan frame, it is possible to get the desired approximation.

    To proceed this way we need to express the Einstein frame Hubble parameter $\tilde H$ through Jordan frame variables. Formally,
\begin{equation}
\label{HEHJ}
\tilde H=\frac{d \ln(\tilde a)}{d \tilde t}=\Omega^{-1}\left(H+\frac{d \ln{\Omega}}{dt}\right)=
\frac{M_{\mathrm{Pl}}}{\sqrt{F}}\left(H+\frac{F_{,\phi}\dot \phi}{2F}\right).
\end{equation}

    The last expression is nothing else, but the variable $Y$ introduced in Eq.~(\ref{Ydef}) without any connections to the Einstein frame.

    Now, assuming that the second term in the brackets is significantly smaller than the first one, one get
from Eq.~(\ref{H2E}) the following approximate equation:
\begin{equation}
H^2(\phi)\approx \frac{V}{3F}
\end{equation}
and
\begin{equation}
\label{chiappr1a}
\chi(\phi)=\frac{\dot\phi H}{H^2}\approx \frac{2F\left(2F_{,\phi}V-FV_{,\phi}\right)}{V\left(2F+3F_{,\phi}^2\right)}.
\end{equation}

    The parameters $\varepsilon_1(\phi)$ and $\zeta_1(\phi)$ are obtained  from Eq.~(\ref{eps1zeta1phi}):
\begin{equation}
\label{eps1appr1a}
\varepsilon_1(\phi)\approx {}-\frac{F\left(2F_{,\phi}V-FV_{,\phi}\right)}{V(2F+3F_{,\phi}^2)}\left(\frac{V_{,\phi}}{V}-\frac{F_{,\phi}}{F}\right),
\end{equation}
\begin{equation}
\label{zeta1appr1a}
\zeta_1(\phi)=\frac{F_{,\phi}}{F}\chi(\phi)\approx \frac{2F_{,\phi}\left(2F_{,\phi}V-FV_{,\phi}\right)}{V\left(2F+3F_{,\phi}^2\right)}\,.
\end{equation}
Expressions for $\varepsilon_2(\phi)$ and $\zeta_2(\phi)$ are given in Appendix A. We will refer to this approximation as the known one.

    Note, however, that when we actually transform from the Einstein frame to the Jordan frame the expression for $\tilde H^2$ contains other terms, and the system under consideration is in fact an {\it additional} approximation of the Einstein frame approximation rather than a direct transformation of the Einstein slow-roll approximation. This motivates us to search more accurate approximations.

\section{New slow-roll approximations}

\subsection{New approximation I}
~~~~Neglecting terms proportional to ${\dot\phi}^2$ and $\ddot{\phi}$, we reduce Eqs.~(\ref{equ00}) and (\ref{KG}) as follows
\begin{equation}
\label{00appr1}
3H^2F\approx V-3F_{,\phi}\dot\phi H\,,
\end{equation}
\begin{equation}
\label{GBappr1}
3H\dot\phi+ V_{,\phi}-3F_{,\phi}(\dot H+2H^2)\approx 0.
\end{equation}

    Differentiating Eq.~(\ref{00appr1}) and expressing $V_{,\phi}$ via Eq.~(\ref{GBappr1}),
we get that $\dot{H}$ is defined by Eq.~(\ref{dotH}).

    Substituting (\ref{00appr1}) and (\ref{dotH}) into Eq.~(\ref{GBappr1}) and again neglecting terms proportional to ${\dot\phi}^2$ and $\ddot{\phi}$, we find
\begin{equation}
\label{dotphiH1}
3H\dot\phi\approx -2\left(\frac{FV_{,\phi}-2F_{,\phi}V}{2F+3F_{,\phi}^2}\right),
\end{equation}

    Substituting (\ref{dotphiH1}) into Eq.~(\ref{00appr1}), we obtain
\begin{equation}
\label{H2appr1}
H^2(\phi)\approx \frac{2FV-F_{,\phi}^2V+2FF_{,\phi}V_{,\phi}}{3F(2F+3F_{,\phi}^2)}\,.
\end{equation}
Therefore,
\begin{equation}
\label{chiappr1}
\chi(\phi)=\frac{\dot\phi H}{H^2}\approx -\frac{2F(FV_{,\phi}-2F_{,\phi}V)}{2FV+2FF_{,\phi}V_{,\phi}-F_{,\phi}^2V}\,.
\end{equation}

    We find the slow-roll parameters $\varepsilon_1(\phi)$ and $\zeta_1(\phi)$, applying Eq.~(\ref{eps1zeta1phi}):
\begin{equation}
\label{eps1appr1}
\begin{split}
\varepsilon_1(\phi)&\approx{}-\frac{F(2F_{,\phi}V-FV_{,\phi})}{2FV+2FF_{,\phi}V_{,\phi}-F_{,\phi}^2V}\\
\\&\times\left(\frac{2FV_{,\phi}+2F_{,\phi}V+F_{,\phi}^2V_{,\phi}+2FF_{,\phi\phi}V_{,\phi}+2FF_{,\phi}V_{,\phi\phi}-2F_{,\phi}F_{,\phi\phi}V}{2FV+2FF_{,\phi}V_{,\phi}-F_{,\phi}^2V}
\right.\\&{}\left.-\frac{F_{,\phi}}{F}-\frac{2F_{,\phi}(1+3F_{,\phi\phi})}{2F+3F_{,\phi}^2}\right)\,,
\end{split}
\end{equation}
\begin{equation}
\label{zeta1appr1}
\zeta_1(\phi)=\frac{F_{,\phi}}{F}\chi(\phi)\approx\frac{ 2F_{,\phi}(2F_{,\phi}V-FV_{,\phi})}{2FV+2FF_{,\phi}V_{,\phi}-F_{,\phi}^2V}.
\end{equation}

It follows from Eqs.~(\ref{H2appr1}) and (\ref{zeta1appr1}) that
\begin{equation}
    H^2(\phi)=\frac{V}{3F\left(1+\zeta_1(\phi)\right)}\,.
\end{equation}

Expressions for $\varepsilon_2(\phi)$ and $\zeta_2(\phi)$ are presented in Appendix~A.

\subsection{New slow-roll approximations II and III}
~~~~These slow-roll approximations are based on system~(\ref{equY})--(\ref{KGequVeff}).
Neglecting the proportional to $\dot{\phi}^2$ term in Eq.~(\ref{equY}), we get
\begin{equation}
\label{Yequappr}
Y^2\approx \frac{V_{\mathrm{eff}}}{3M_{\mathrm{Pl}}^2}=\frac{M_{\mathrm{Pl}}^2V}{3F^2}\,.
\end{equation}

    Differentiating this equation over time and using Eq.~(\ref{Fr21Qm}), we obtain
\begin{equation}
\label{phiequappr}
\psi\approx{}-\frac{M_{\mathrm{Pl}}V_{\mathrm{eff},\phi}}{3YA\sqrt{F}}= {}-\frac{2M_{\mathrm{Pl}}\left(FV_{,\phi}-2F_{,\phi}V\right)}{3Y\sqrt{F}\left(2F+3F_{,\phi}^2\right)}
    = {}-\frac{2\left(FV_{,\phi}-2F_{,\phi}V\right)}{3H\left(1+\frac{\zeta_1}{2}\right)\left(2F+3F_{,\phi}^2\right)}\,.
\end{equation}

    So, the system of equations (\ref{Fr21Qm}), (\ref{Yequappr}), and (\ref{phiequappr}) is self-consistent. Note that Eq.~(\ref{phiequappr}) can be obtained from Eq.~(\ref{KGequVeff}) if we neglect terms proportional to $\ddot{\phi}$ and $\dot{\phi}^2$.

    Using Eqs.~(\ref{Yzeta}) and (\ref{Yequappr}), we get from Eq.~(\ref{phiequappr}):
\begin{equation}
\label{dphiY}
\chi\approx{}-\frac{2\left(FV_{,\phi}-2F_{,\phi}V\right)}{3H^2\left(1+\frac{\zeta_1}{2}\right)\left(2F+3F_{,\phi}^2\right)}
 ={}-\frac{2F\left(FV_{,\phi}-2F_{,\phi}V\right)}{V\left(2F+3F_{,\phi}^2\right)}\left(1+\frac{F_{,\phi}}{2F}\chi\right)\,.
\end{equation}

    Therefore, we obtain the first-order differential equation that defines slow-roll dynamic of $\phi$:
\begin{equation}
\label{equphiNslr}
\chi=\frac{d\phi}{dN}=\frac{2F\left(2F_{,\phi}V-FV_{,\phi}\right)}{2FV+F_{,\phi}FV_{,\phi}+F_{,\phi}^2V}\,.
\end{equation}

    Substituting (\ref{equphiNslr}) into Eq.~(\ref{eps1zeta1phi}), we find
    \begin{equation}
\label{zeta1appr23}
\zeta_1(\phi)=\frac{2F_{,\phi}(2F_{,\phi}V-FV_{,\phi})}{2FV+FF_{,\phi}V_{,\phi}+F_{,\phi}^2V}\,.
\end{equation}

    If we neglect the term proportional to $\zeta_1^2$ in
\begin{equation}
\label{H2zeta1}
H^2=\frac{FY^2}{M_{\mathrm{Pl}}^2\left(1+\frac{1}{2}\zeta_1\right)^2}\approx\frac{V}{3F\left(1+\zeta_1\right)},
\end{equation}
then we get
\begin{equation}
\label{H2appr2}
H^2(\phi)\approx\frac{V}{3F(1+\zeta_1)}=\frac{V\left(2FV+FF_{,\phi}V_{,\phi}+F_{,\phi}^2V\right)}{3F\left(2FV-FF_{,\phi}V_{,\phi}+5F_{,\phi}^2V\right)}.
\end{equation}
So,
\begin{equation}
\label{eps1appr2}
\begin{split}
\varepsilon_1(\phi)&\approx \frac{\zeta_1}{2}\left(1-\frac{FV_{,\phi}}{VF_{,\phi}}+\frac{F}{F_{,\phi}}\,\frac{d\ln(\zeta_1+1)}{d\phi}\right)={} -\frac{F(2F_{,\phi}V-FV_{,\phi})}{2FV+FF_{,\phi}V_{,\phi}+F_{,\phi}^2V}
\\&\times\left(\frac{V_{,\phi}}{V}+\frac{2FV_{,\phi}+2F_{,\phi}V+2F_{,\phi}^2V_{,\phi}+FF_{,\phi\phi}V_{,\phi}+FF_{,\phi}V_{,\phi\phi}+2F_{,\phi}F_{,\phi\phi}V}{2FV+FF_{,\phi}V_{,\phi}+F_{,\phi}^2V}\right.
\\&\left.{} -\frac{F_{,\phi}}{F}-\frac{2FV_{,\phi}+2F_{,\phi}V+4F_{,\phi}^2V_{,\phi}-FF_{,\phi\phi}V_{,\phi}-FF_{,\phi}V_{,\phi\phi}+10F_{,\phi}F_{,\phi\phi}V}{2FV-FF_{,\phi}V_{,\phi}+5F_{,\phi}^2V}\right).
\end{split}
\end{equation}
We call this approximation as the approximation II.
Expressions for $\varepsilon_2(\phi)$ and $\zeta_2(\phi)$ are presented in Appendix A. In Appendix B, we show how the approximation II can be obtained without the use of system~(\ref{equY})--(\ref{KGequVeff}).

    If we do not neglect any terms in Eq.~(\ref{H2zeta1}) and use Eq.~(\ref{equphiNslr}), then we obtain other approximate formulae for $H^2(\phi)$ and $\varepsilon_1(\phi)$:
\begin{equation}
\label{H2appr3}
H^2\approx\frac{V}{3F}\left(1+\frac{F_{,\phi}\chi}{2F}\right)^{-2}=\frac{\left(2FV+FF_{,\phi}V_{,\phi}+F_{,\phi}^2V\right)^2}{3FV\left(2F+3F_{,\phi}^2\right)^2}\,,
\end{equation}

\begin{equation}
\begin{array}{l}
\label{eps1appr3}
\varepsilon_1(\phi)={}-\frac{\zeta_1F}{2F_{,\phi}}\left(\frac{V_{,\phi}}{V}-\frac{F_{,\phi}}{F}-2\frac{d\ln\left(1+\frac{\zeta_1}{2}\right)}{d\phi}\right)={} -\frac{F(2F_{,\phi}V-FV_{,\phi})}{2FV+FF_{,\phi}V_{,\phi}+F_{,\phi}^2V}\times\\
\\\left[\frac{2\left(2FV_{,\phi}+2F_{,\phi}V+2F_{,\phi}^2V_{,\phi}+FF_{,\phi\phi}V_{,\phi}+FF_{,\phi}V_{,\phi\phi}+2F_{,\phi}F_{,\phi\phi}V\right)}{2FV+FF_{,\phi}V_{,\phi}+F_{,\phi}^2V}
-\frac{V_{,\phi}}{V}-\frac{F_{,\phi}}{F}-\frac{4F_{,\phi}\left(1+3F_{,\phi\phi}\right)}{2F+3F_{,\phi}^2}\right].
\end{array}
\end{equation}

     In fact, this is the approximation that fully corresponds to the Einstein frame approximation. Indeed, all terms in Eq.~(\ref{equY}) represent terms from the corresponding Friedmann equation in the Einstein frame. As we have already mentioned, the function $Y$ is the Hubble parameter in the Einstein frame, so Eq.~(\ref{equY}) differs from the standard form of the Friedmann equation only because there is no transition to the Einstein frame scalar field, so that the kinetic term in Eq.~(\ref{equY}) is a non-standard one and the time $t$ is a parametric one in the Einstein frame. It can be easily checked that this term transforms to the standard kinetic term after appropriate field and time redefinitions. So, neglecting the kinetic term and keeping all other terms, we start with the slow-roll approximation in the Einstein frame and return to the Jordan frame without further simplifications. This approximation is referred to as the approximation III.

     Using Eq.~(\ref{HEHJ}), we express the Einstein frame slow-roll parameter $\tilde{\varepsilon}_1$ via the Jordan frame slow-roll parameters as follows:
\begin{equation}
\tilde{\varepsilon}_1={}-\frac{1}{\tilde{H}^2}\,\frac{d\tilde{H}}{d\tilde{t}}=\varepsilon_1+\frac{\zeta_1\left(1-\varepsilon_1\right)}{2+\zeta_1}+\frac{2\zeta_1\zeta_2}{\left(2+\zeta_1\right)^2}\approx \varepsilon_1+\frac12\zeta_1\,.
\end{equation}
So, the moments of the end of inflation in the Jordan and Einstein frames do not coincide. This means that despite the above-mentioned correspondence, calculations in this approximation give results, in principle, different from those obtained directly in the Einstein frame.

\section{The Higgs-driven inflationary model}
~~~~Let us consider the well-known inflationary model~\cite{Spokoiny:1984bd,Makino:1991sg,Barvinsky:1994hx,Bezrukov:2007ep,Bezrukov:2013fka} with the induced gravity term and the fourth-degree monomial potential,
\begin{equation}
\label{phi4model}
F(\phi)= M_{\mathrm{Pl}}^2+\xi\phi^2,\qquad V(\phi)=\frac{\lambda}{4}\phi^4,
\end{equation}
where $\xi$ and $\lambda$ are positive constants. Note that the function $\phi(N)$, a solution of Eq.~(\ref{PhiSYSN}), as well as functions $\chi(\phi)$, $\varepsilon_i(\phi)$ and $\zeta_i(\phi)$ calculated in any slow-roll approximation do not depend on $\lambda$. Hence, the values of inflationary parameters $n_s$ and $r$ do not depend on $\lambda$ and this parameter is important to define the inflationary parameter $A_s$ only.

In Refs.~\cite{Akin:2020mcr,Jarv:2021qpp}, this model has been used to compare two known slow-roll approximations mentioned in Section 4. It has been shown by calculation of the inflationary parameters, which are to be compared with the observational data, and by phase portrait construction that the second approach is essentially more accurate than the first one. Our calculations confirm this result. In figures, we compare the proposed approximations
only with the more accurate known approximation, described in Subsection 4.2.

\begin{figure}
\includegraphics[scale=0.25]{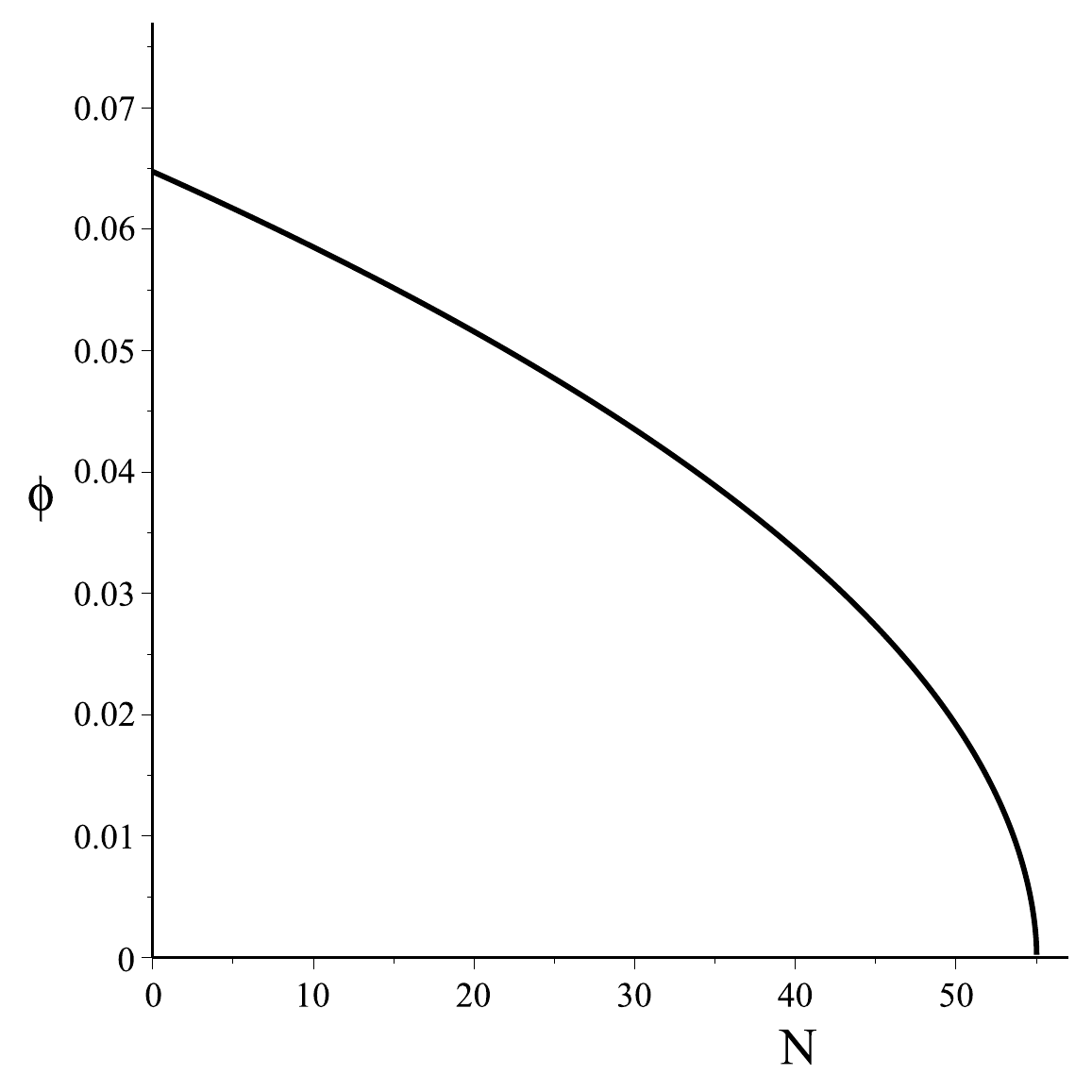}
\includegraphics[scale=0.27]{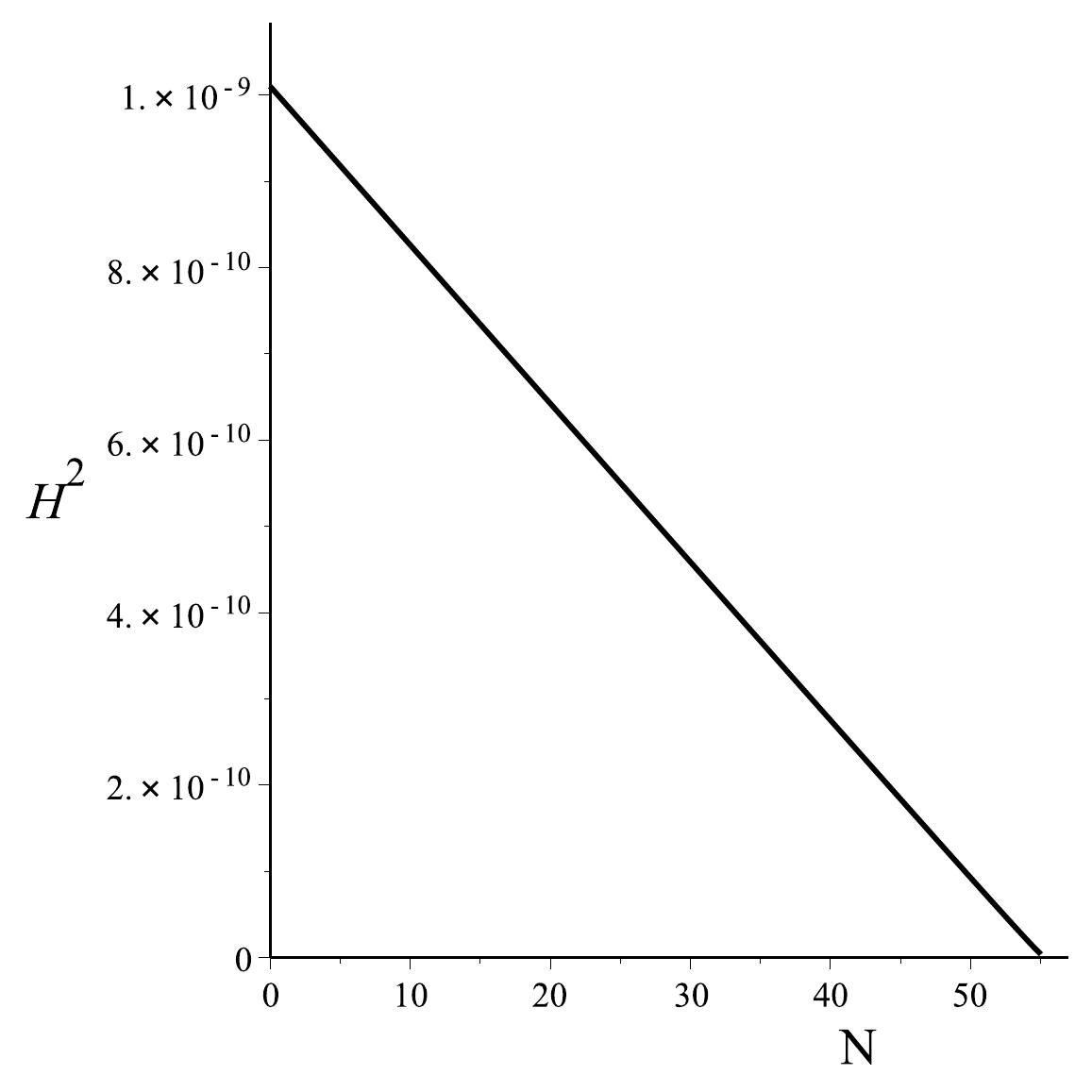}
\includegraphics[scale=0.25]{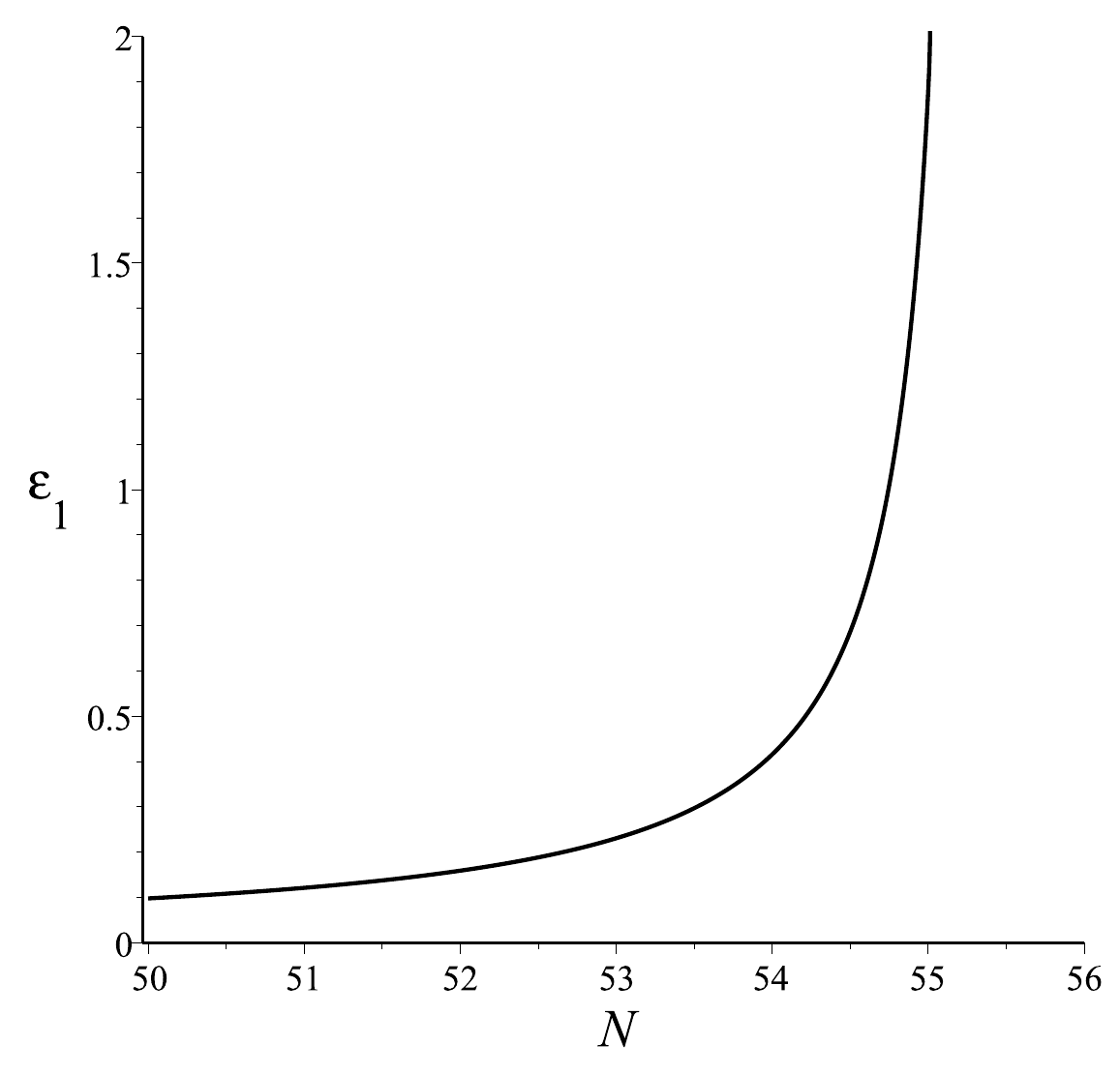}
\caption{The functions $\phi(N)$ (left), $H^2(N)$ (center), and $\varepsilon_1(N)$ (right) for the model with the potential $V(\phi)=0.0125\phi^4$ and the coupling function $F(\phi)=1+17367\phi^2$. The curves are obtained by the numerical integration of the system (\ref{PhiSYSN}) with the initial data $\phi(N_0)=0.077$, $\chi(N_0)=0$} at $N_0=-23$.
\label{FigHiggsN}
\end{figure}

   With an additional assumption that $\phi$ is the Standard model Higgs boson, this model is known as the Higgs-driven inflationary model~\cite{Bezrukov:2007ep} (see also
Refs.~\cite{Barvinsky:2008ia,Garcia-Bellido:2008ycs,DeSimone:2008ei,Bezrukov:2010jz,Bezrukov:2013fka,Bezrukov:2014ipa,Ren:2014sya,vandeBruck:2015gjd}). Following~\cite{vandeBruck:2015gjd}, we choose $\xi=17367$
and $\lambda=0.05$ for the Higgs-driven inflationary model. In Fig.~\ref{FigHiggsN}, we demonstrate the evolution of $\phi(N)$ and $H^2(N)$ during inflation. Values of $\phi$ and $H$ are given in units of $M_{\mathrm{Pl}}$ in all figures. The value of $\phi$, at which the inflationary parameters are calculated, corresponds to $N=0$.  The right picture of Fig.~\ref{FigHiggsN} shows that the inflation ends at $N\approx 55$.

\begin{figure}
\includegraphics[scale=0.39]{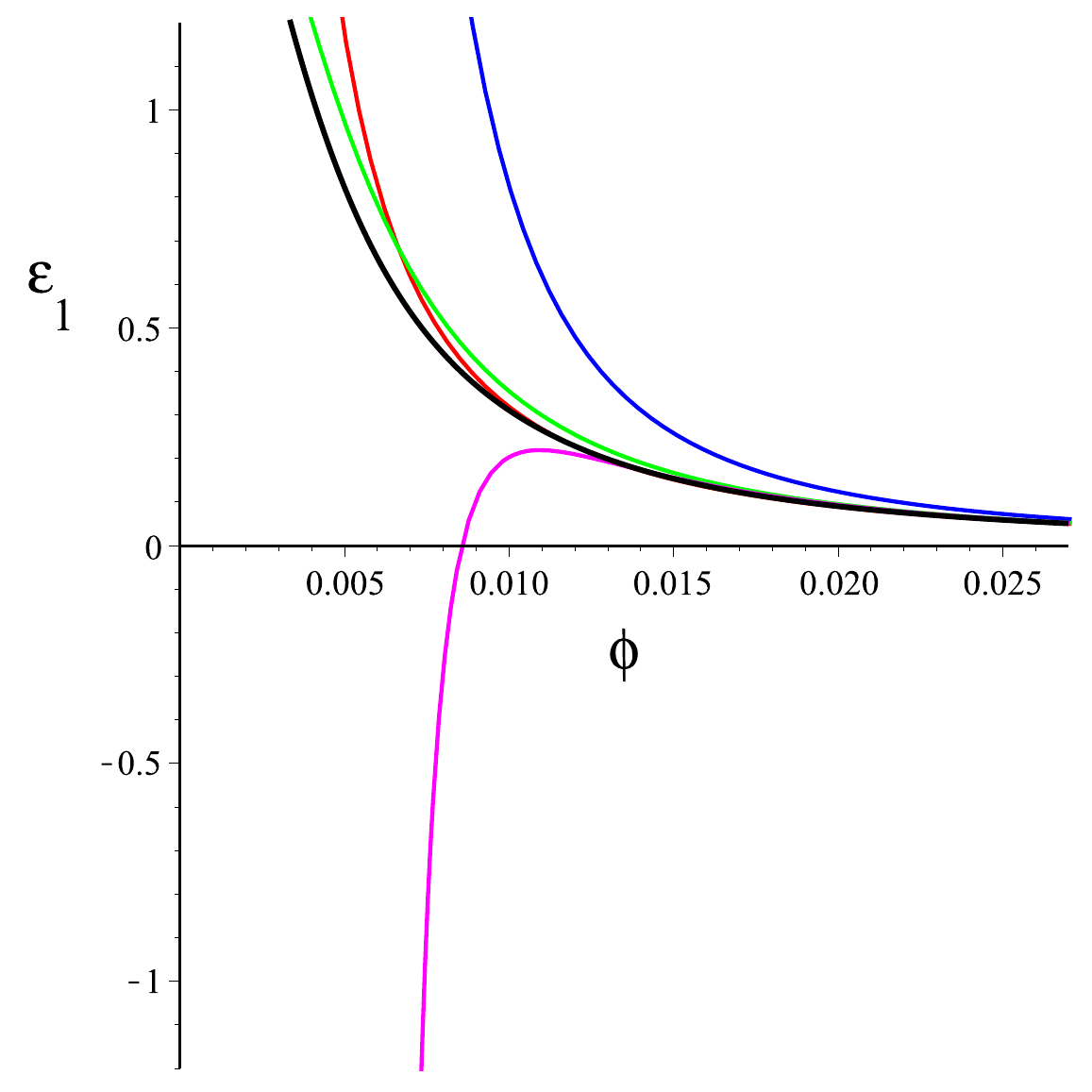}\quad
\includegraphics[scale=0.39]{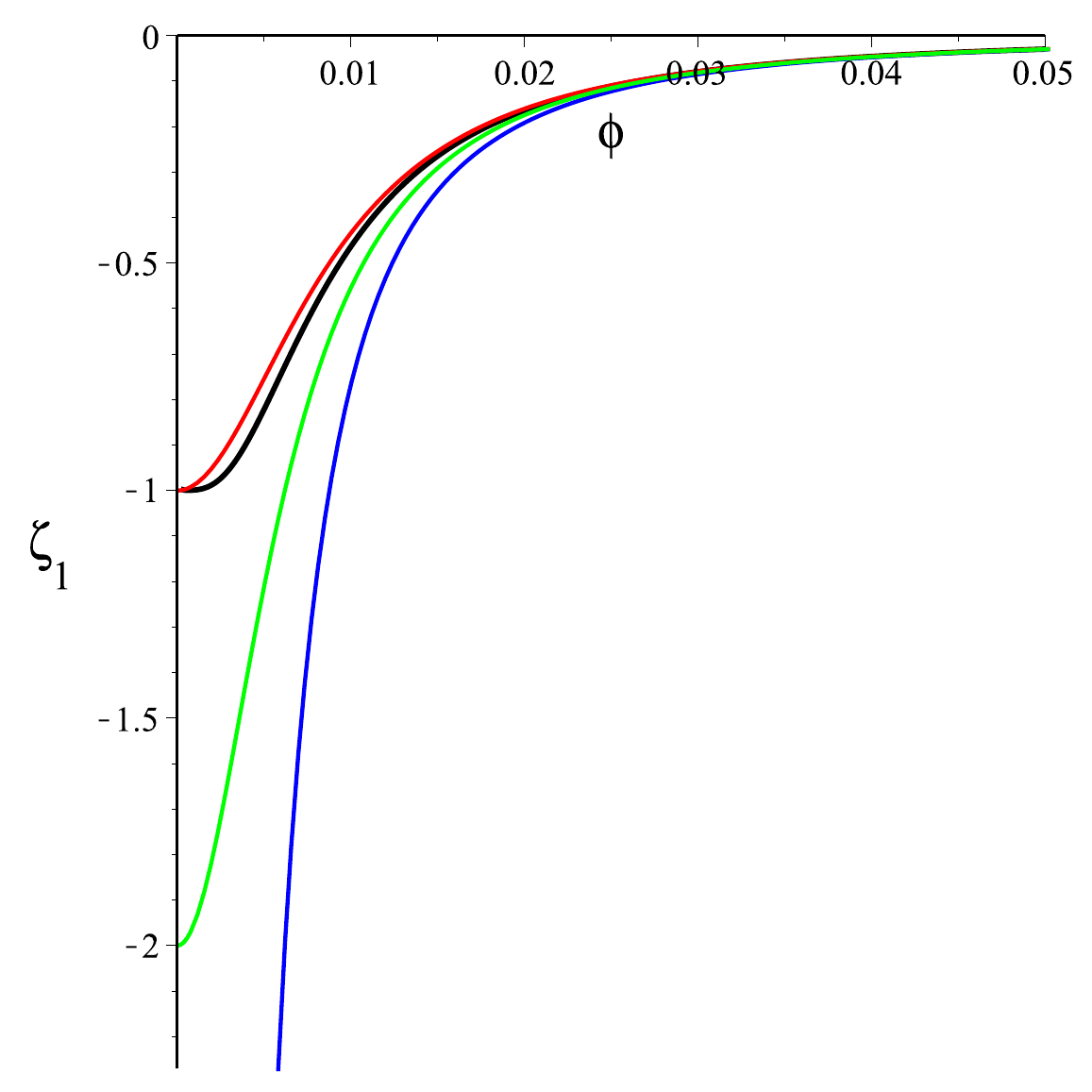}
\caption{The functions $\varepsilon_1(\phi)$ (left) and $\zeta_1(\phi)$ (right) for the model with the potential $V(\phi)=0.0125\phi^4$ and the coupling function $F(\phi)=1+17367\phi^2$. The black lines are the result of the numerical integration of the system (\ref{PhiSYSN}), the blue curves are obtained in the known approximation, the red curves are obtained in the approximation I,
the magenta curve is obtained in the approximation II and the green curves are obtained in the approximation III.}
\label{FigHiggs1}
\end{figure}

    In Fig.~\ref{FigHiggs1}, one can see the behavior of the slow-roll parameters $\varepsilon_1$ and $\zeta_1$ as functions of $\phi$. In the left picture, one can see that the value of $\varepsilon_1$ becomes less than $-1$ during inflation in the approximation II. It means that this new approximation cannot be considered as an improved version of the known approximation. On the other hand, the curves that correspond to new approximations I (red) and III (green) are essentially closer to results of numerical integration (the black curve), than the known slow-roll approximation (the blue curve) at the end of inflation.

\begin{figure}
\includegraphics[scale=0.41]{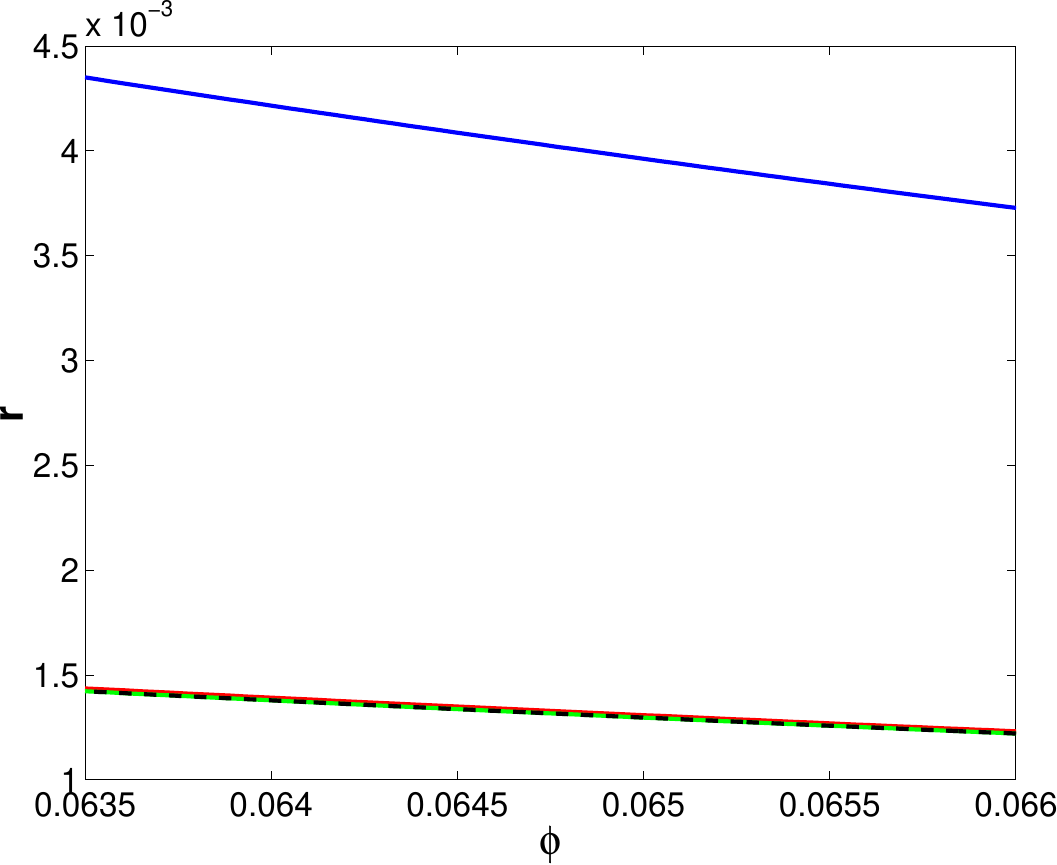}\quad
\includegraphics[scale=0.41]{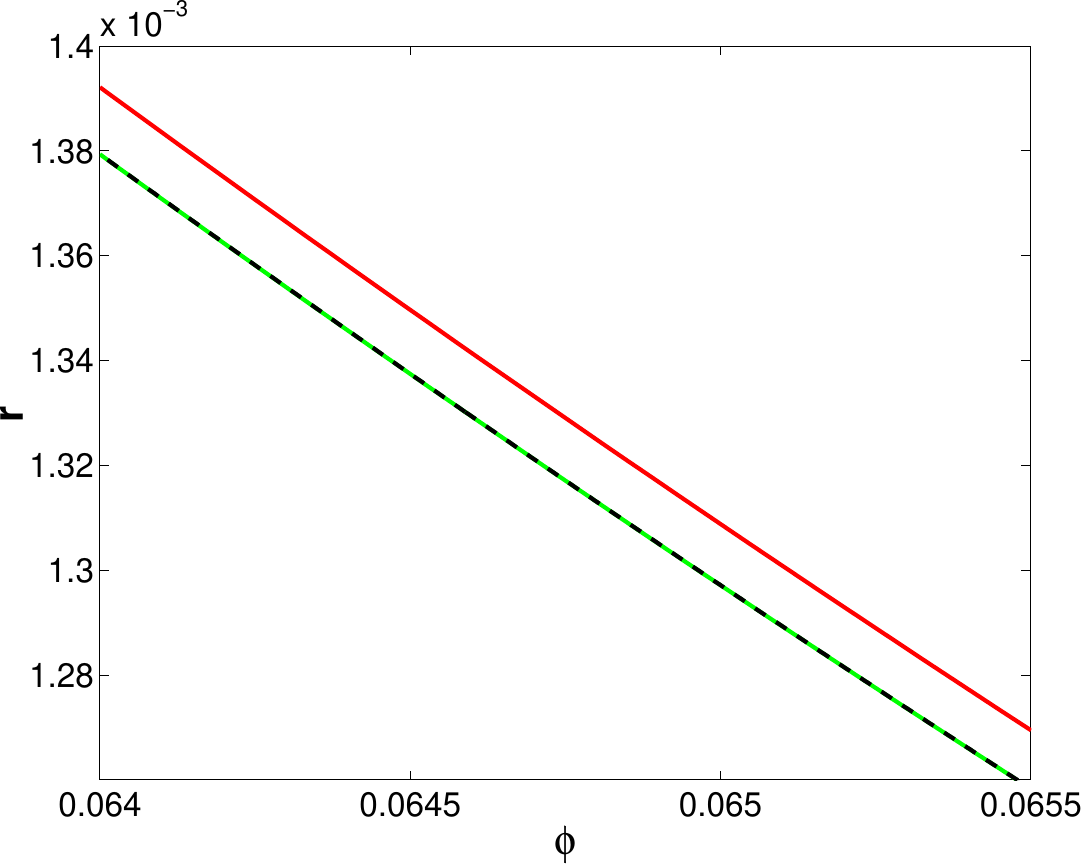}
\caption{The function $r(\phi)$ for the model with the potential $V(\phi)=0.0125\phi^4$ and the coupling function $F(\phi)=1+17367\phi^2$. The black lines are the result of the numerical integration of the system (\ref{PhiSYSN}), the blue curves are obtained in the known approximation, the red curves are obtained in the approximation I and the green curves are obtained in the approximation III.}
\label{FigHiggs2}
\end{figure}

\begin{figure}
\includegraphics[scale=0.27]{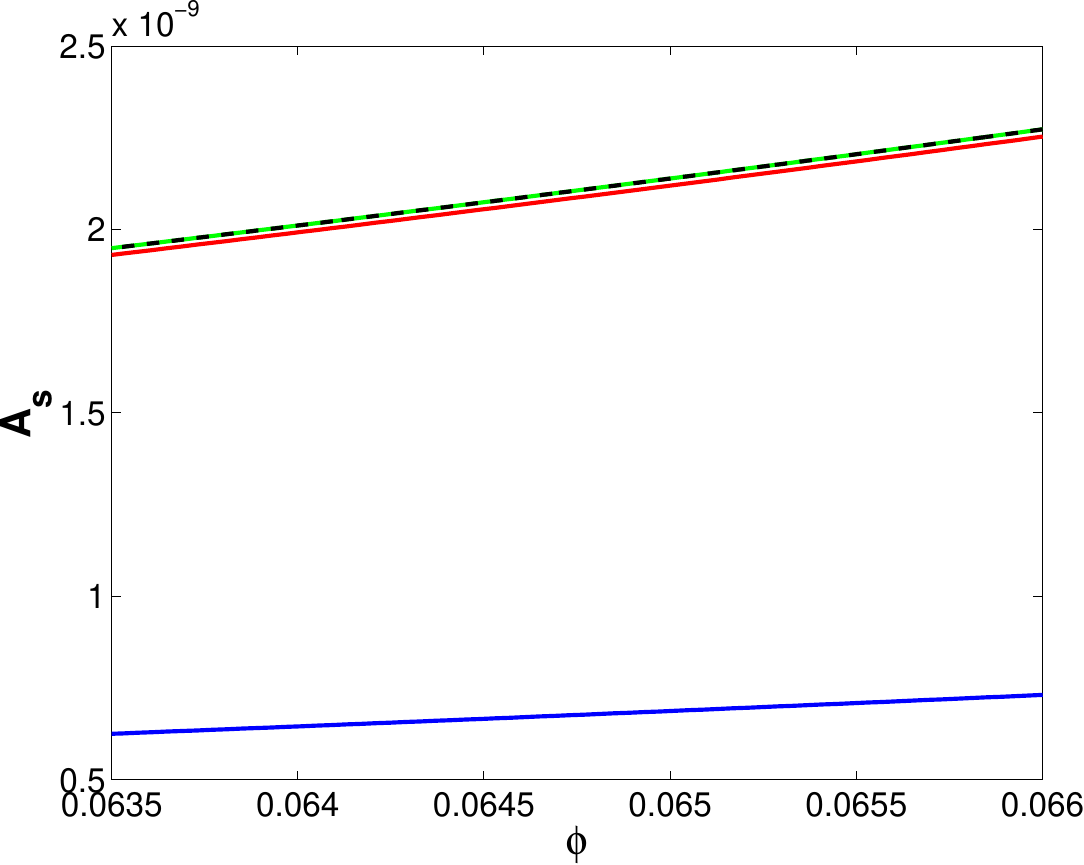}
\includegraphics[scale=0.27]{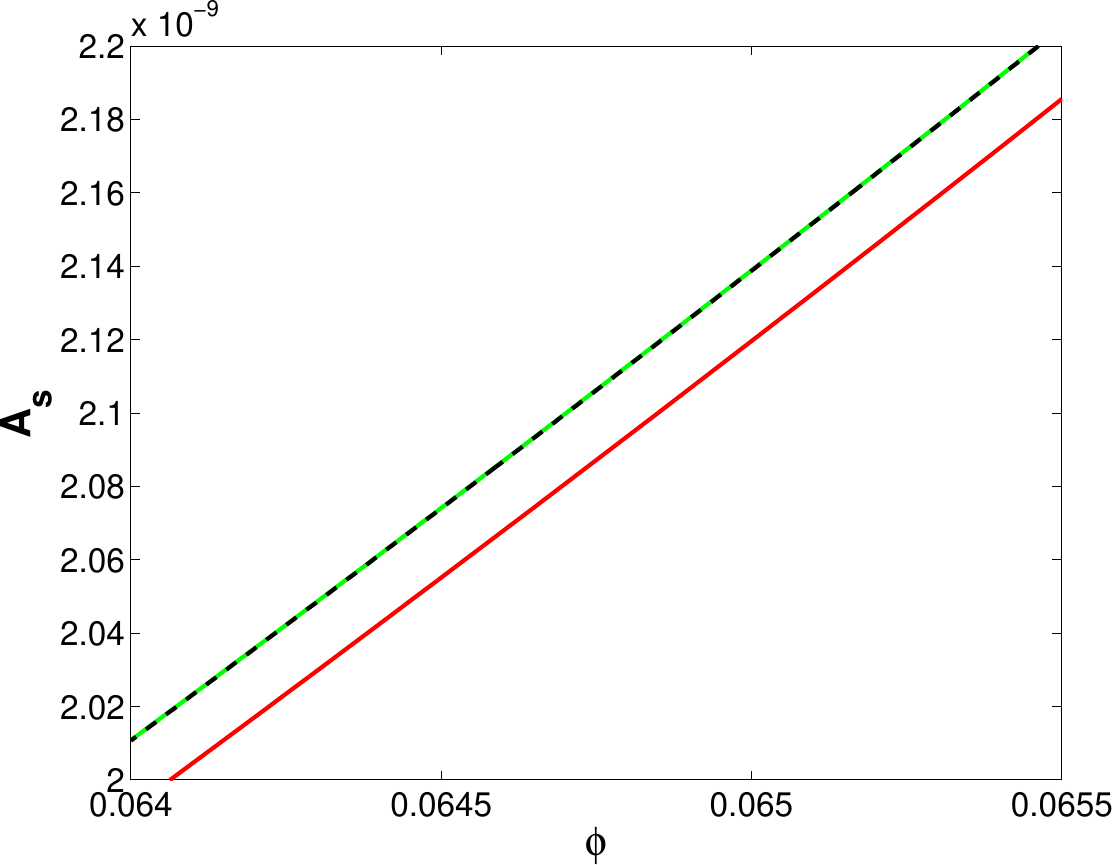}
\includegraphics[scale=0.27]{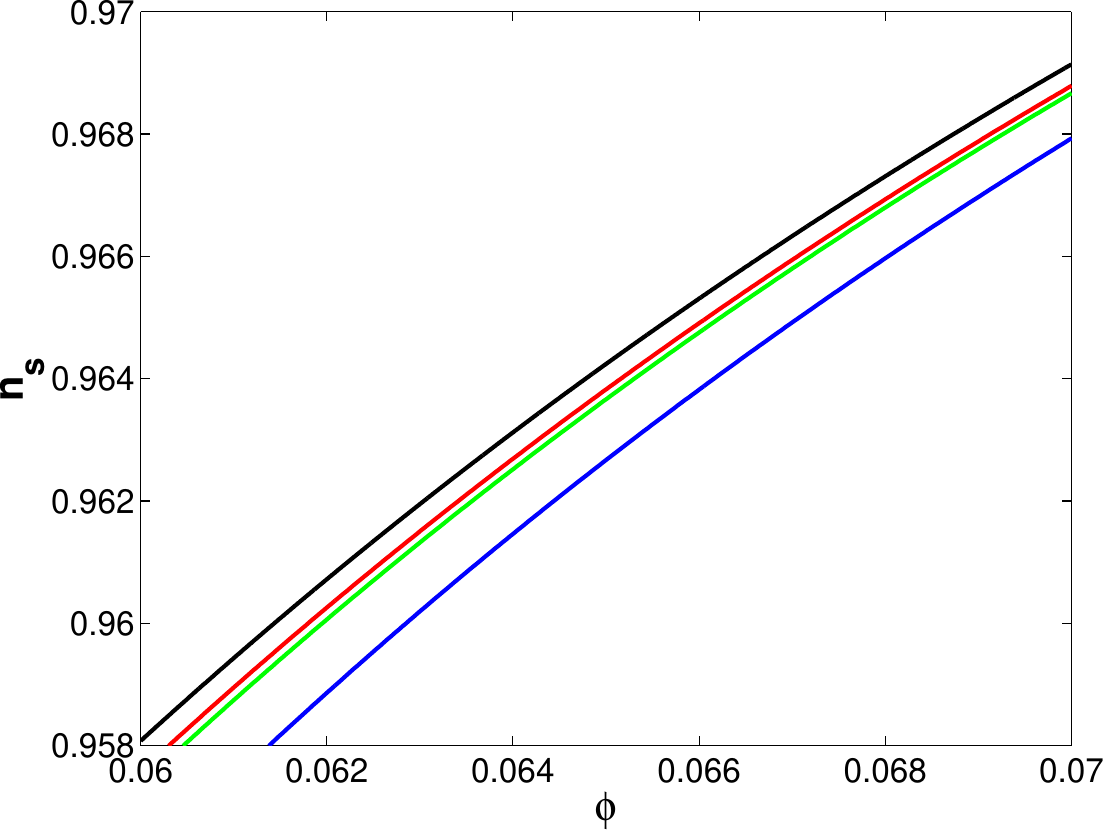}
\caption{The functions $A_s(\phi)$ (the left and center pictures) and $n_s(\phi)$ (the right picture) for the model with the potential $V(\phi)=0.0125\phi^4$ and the coupling function $F(\phi)=1+17367\phi^2$. The black lines are the result of the numerical integration of the system (\ref{PhiSYSN}), the blue curves are obtained in the known approximation, the red curves are obtained in the approximation I, and the green curves are obtained in the approximation III.}
\label{FigHiggs3}
\end{figure}

    A more interesting and unexpected result is that new approximations I and III give essentially more precise values of the inflationary parameters $r$ and $A_s$. One can see in Fig.~\ref{FigHiggs2} that the value of $r$ calculated in the known approximation (the blue curve) differ more than two times from the numerical results (the black curve), whereas new approximations (the red and green curves) give essentially better values of $r$. An incorrect value of $r$ gives an incorrect value of $A_s$ [see Eq.~(\ref{As})] that one can see in Fig.~\ref{FigHiggs3}. On other hand, there are no important difference in values of $n_s$ calculated in the known and new approximations. All of them are sufficiently close to the numerical results (the right picture of Fig.~\ref{FigHiggs3}). So, we come to the conclusion that new approximations I and III are essentially better than the known ones. Note that the approximation III works slightly better than the approximation I for the Higgs-driven inflation.

\begin{figure}
\includegraphics[scale=0.41]{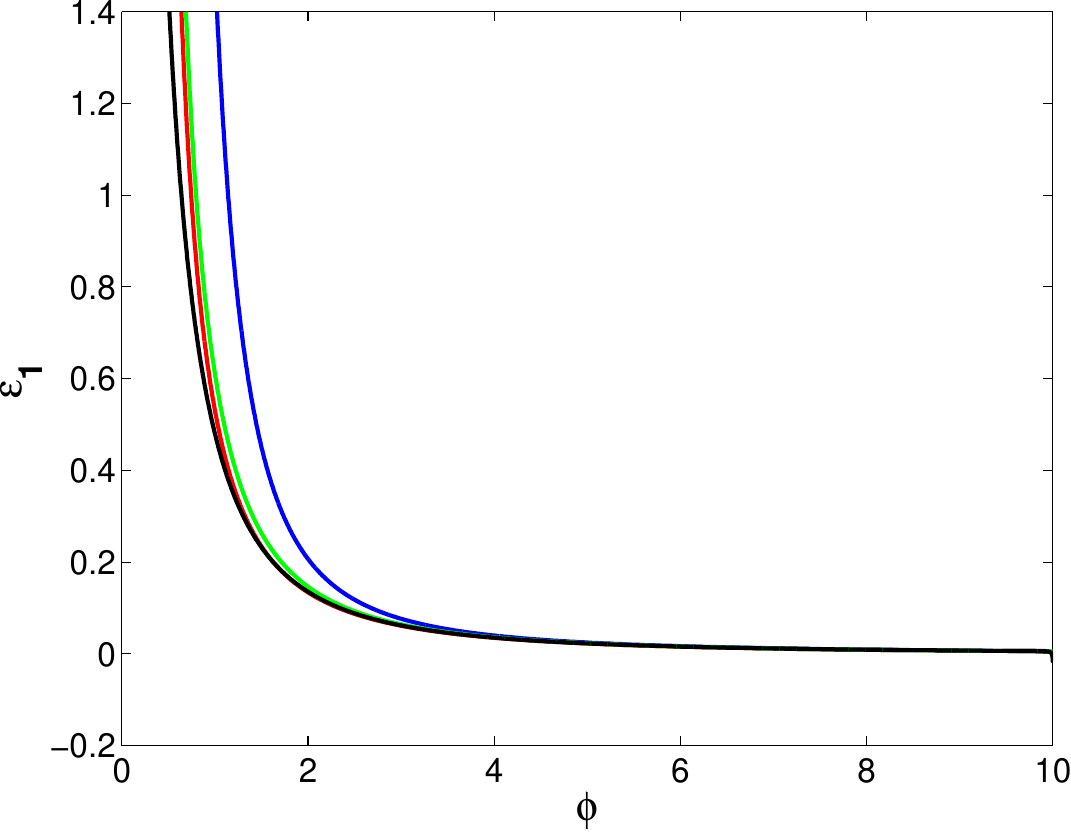}\quad
\includegraphics[scale=0.41]{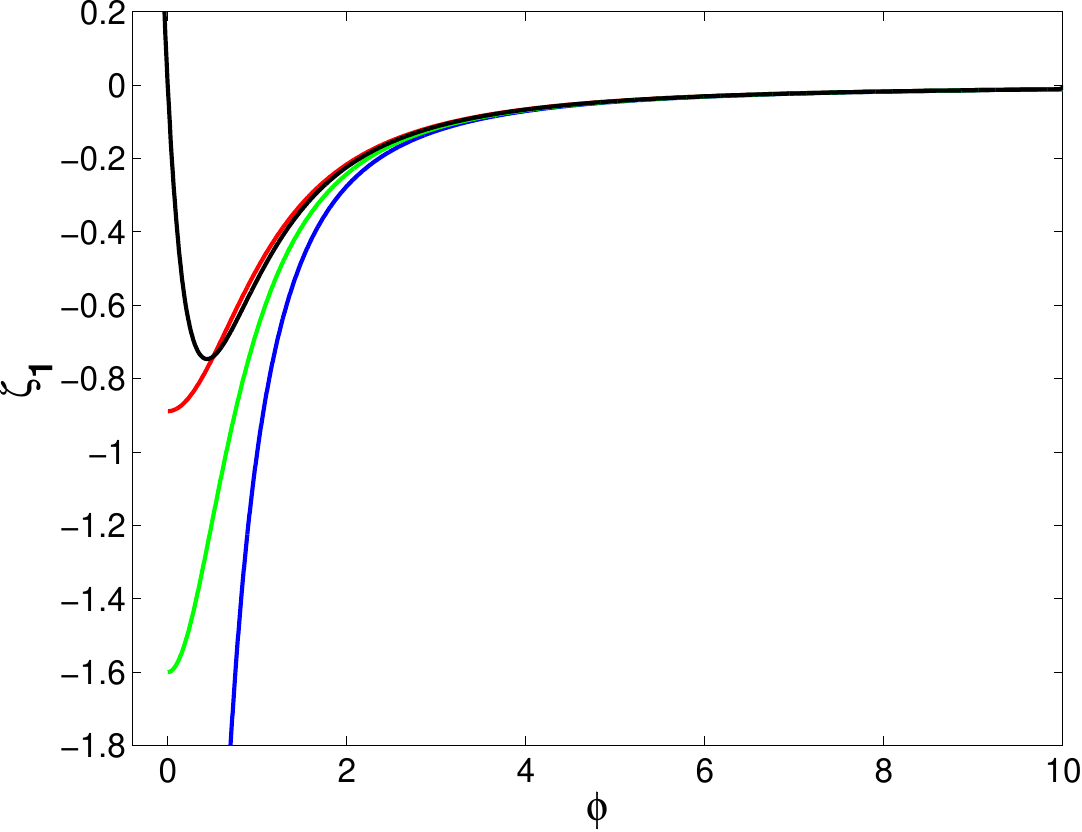}
\caption{The functions $\varepsilon_1(\phi)$ (left) and $\zeta_1(\phi)$ (right) for the model with the potential $V(\phi)=5\cdot 10^{-11}\phi^4$ and the coupling function $F(\phi)=1+\phi^2$. The black lines are the result of the numerical integration of the system (\ref{PhiSYSN}) with the initial data $\phi(N_0)=10$, $\chi(N_0)=0$ at $N_0=-27$, the blue curves are obtained in the known approximation, the red curves are obtained in the approximation I, and the green curves are obtained in the approximation III.}
\label{Fig1}
\end{figure}

\begin{figure}
\includegraphics[scale=0.41]{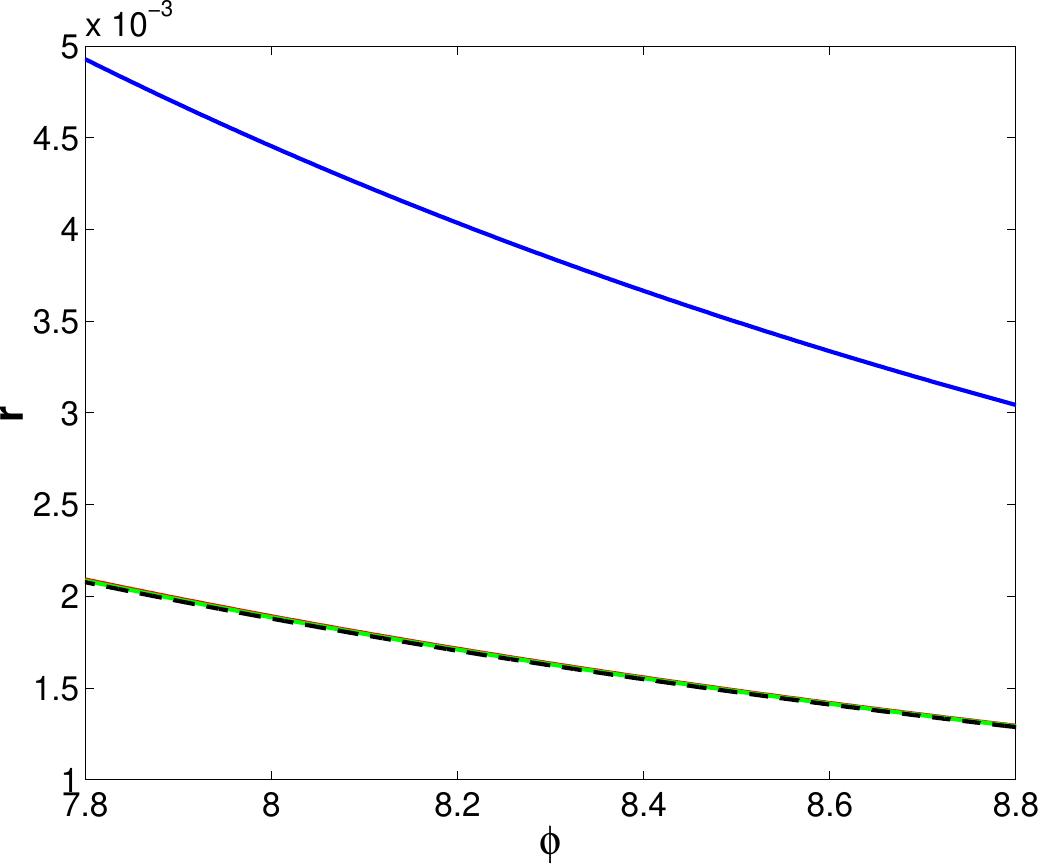}\quad
\includegraphics[scale=0.41]{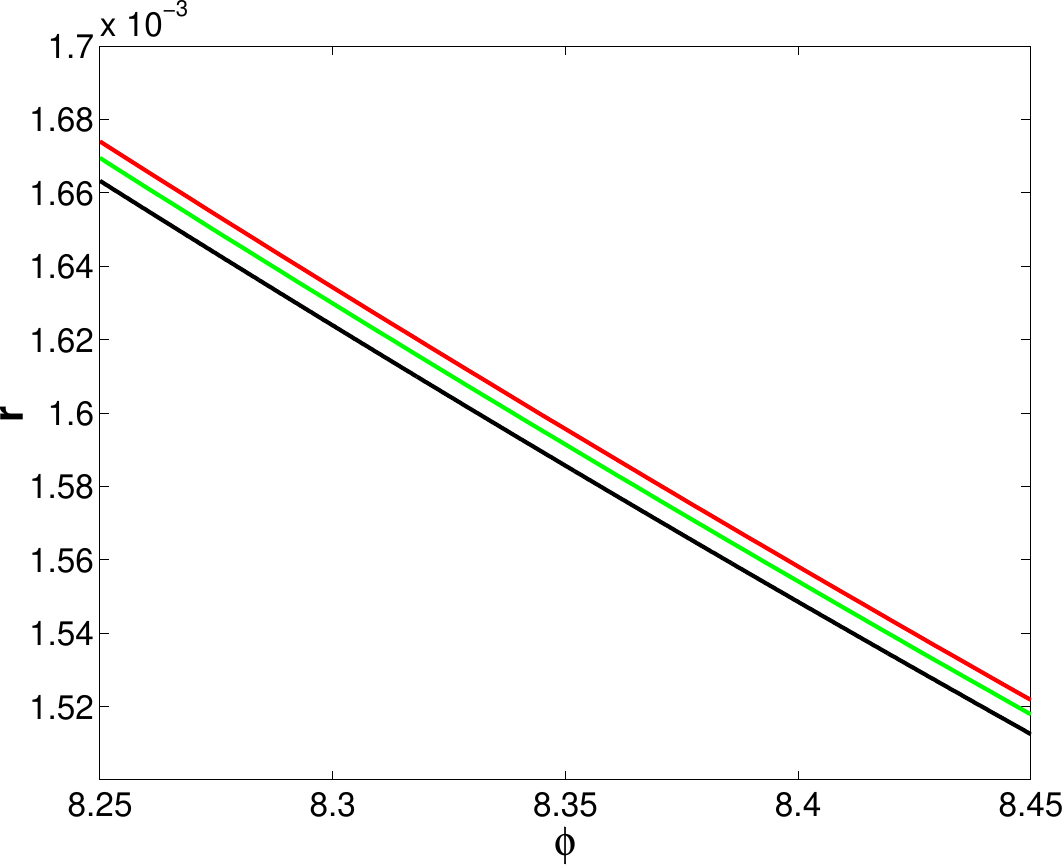}
\caption{The function $r(\phi)$  for the model with the potential $V(\phi)=5\cdot 10^{-11}\phi^4$ and the coupling function $F(\phi)=1+\phi^2$. The black lines are the result of the numerical integration of the system (\ref{PhiSYSN}), the blue curves are obtained in the known approximation, the red curves are obtained in the approximation I, and the green curves are obtained in the approximation III.}
\label{Fig2}
\end{figure}

\begin{figure}
\includegraphics[scale=0.27]{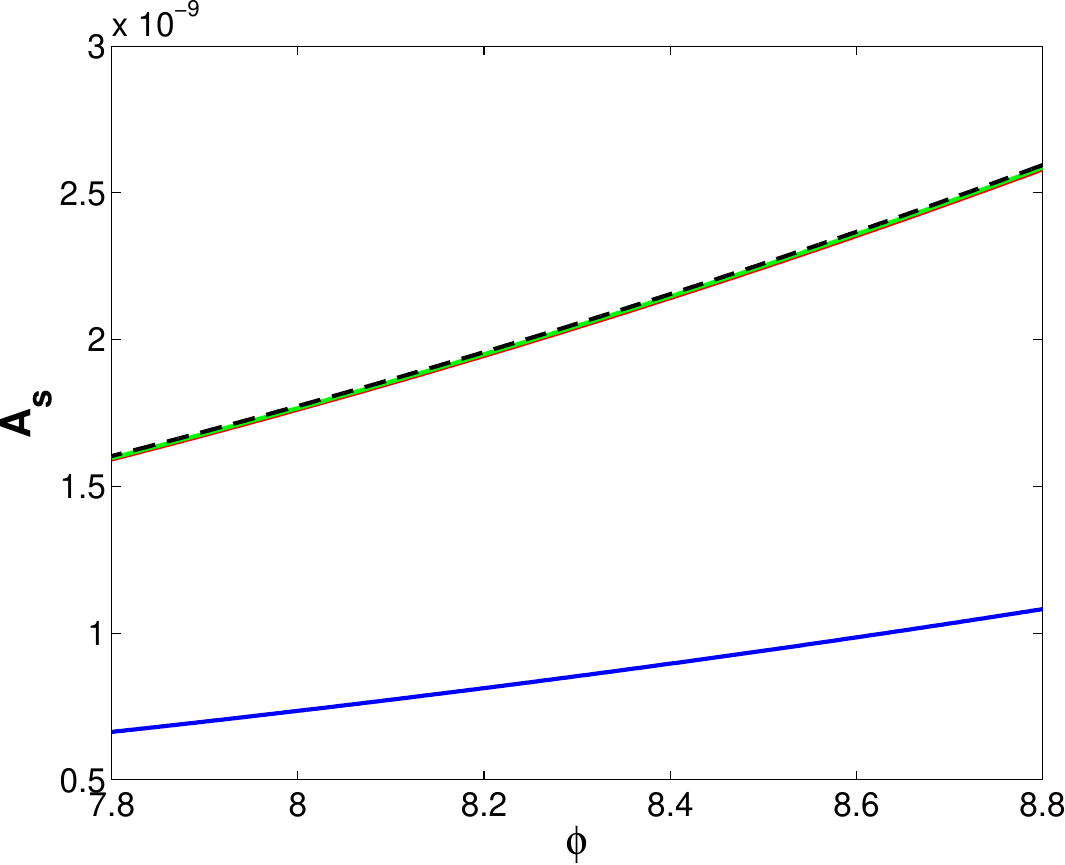}
\includegraphics[scale=0.27]{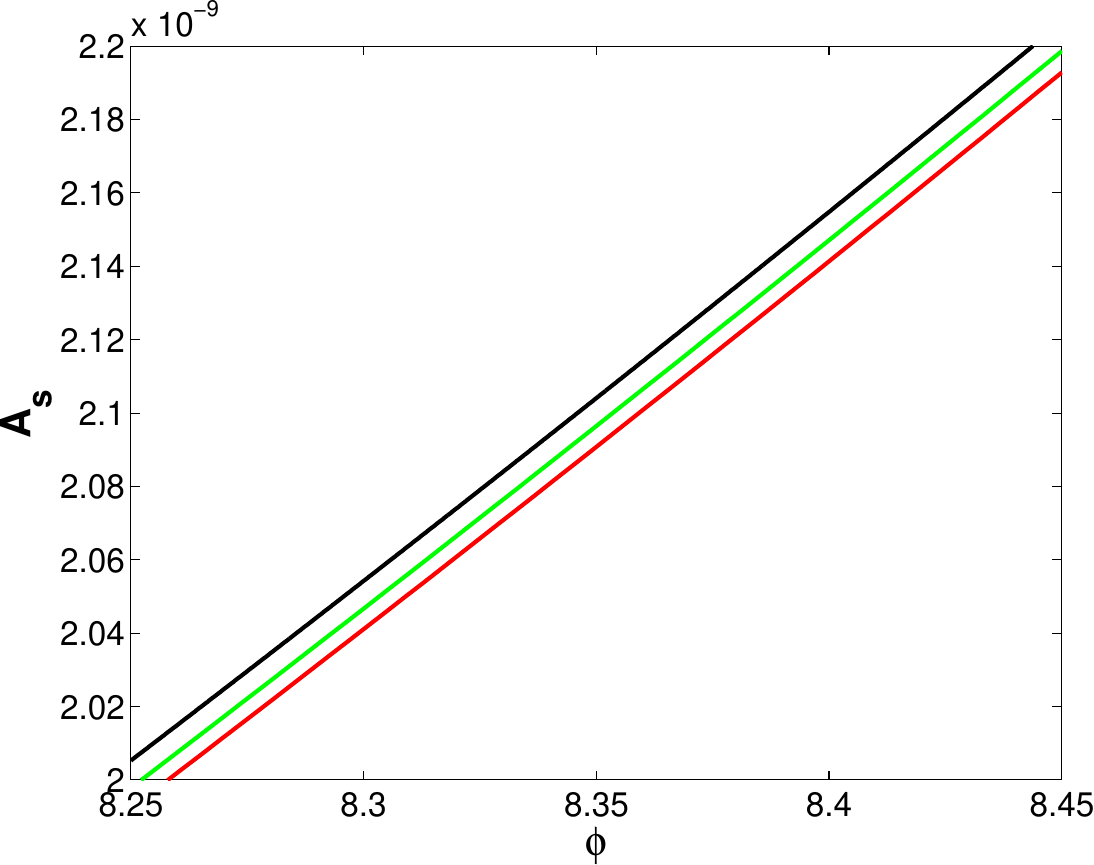}
\includegraphics[scale=0.27]{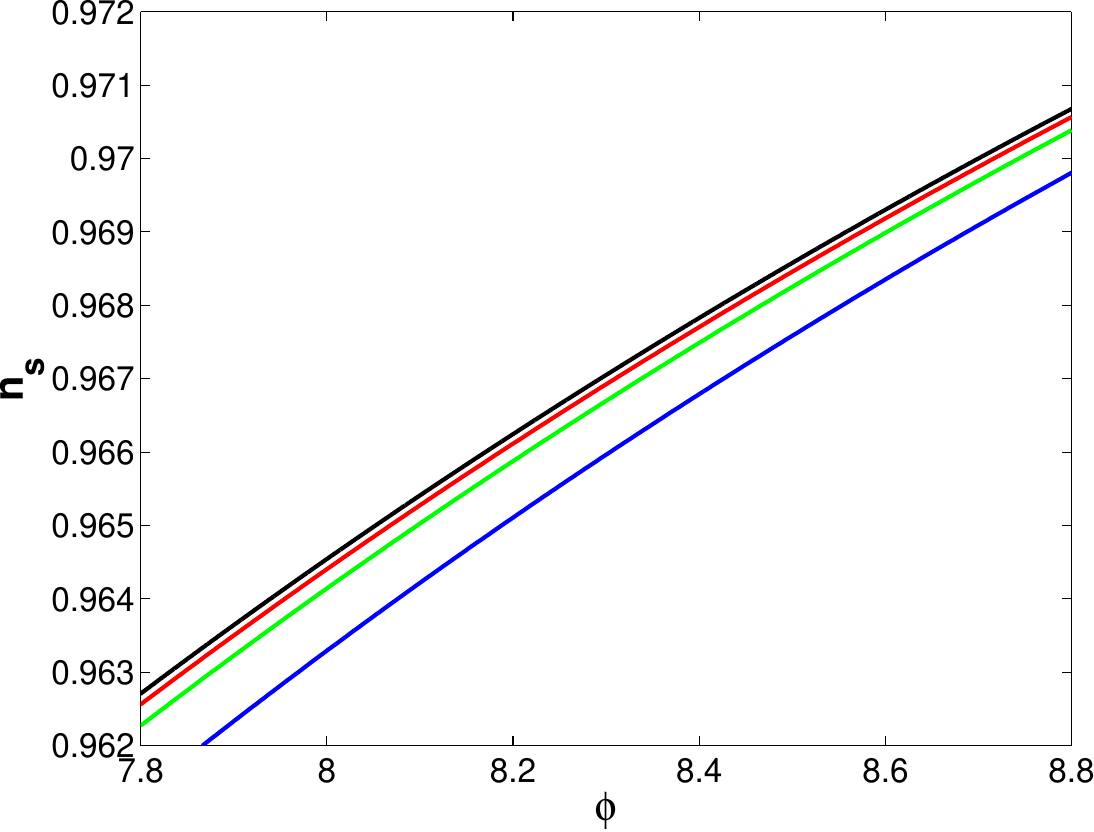}
\caption{The functions $A_s(\phi)$ (the left and center pictures) and $n_s(\phi)$ (the right picture)  for the model with the potential $V(\phi)=5\cdot 10^{-11}\phi^4$ and the coupling function $F(\phi)=1+\phi^2$. The black lines are the result of the numerical integration of the system (\ref{PhiSYSN}), the blue curves are obtained in the known approximation, the red curves are obtained in the approximation I, and the green curves are obtained in the approximation III.}
\label{Fig3}
\end{figure}

    For values of $\xi$ and $\lambda$ that are significantly different from those corresponding to the Higgs-driven inflationary model, the model (\ref{phi4model}) describes inflation as well. In Figs.~\ref{Fig1}--\ref{Fig3}, we compare exact and approximate values of the slow-roll and inflationary parameters for the model (\ref{phi4model}) with $\xi=1$ and $\lambda=2\cdot 10^{-10}$. One can see that curves that correspond to new approximations I (the red curve) and III (the green curve) are essentially closer to results of numerical integration (the black curve), than the known slow-roll approximation (the blue curve) at the end of inflation (Fig.~\ref{Fig1}). For $\xi=1$, new approximations give more precise values of the inflationary parameters $r$ and $A_s$ as in the previous case. One can see this in Figs.~\ref{Fig2} and~\ref{Fig3}. Again, there is no significant difference in values of $n_s$ calculated in the known and new approximations. Note that the approximation I works slightly better than the approximation III at $\xi=1$.

\begin{figure}
\includegraphics[scale=0.41]{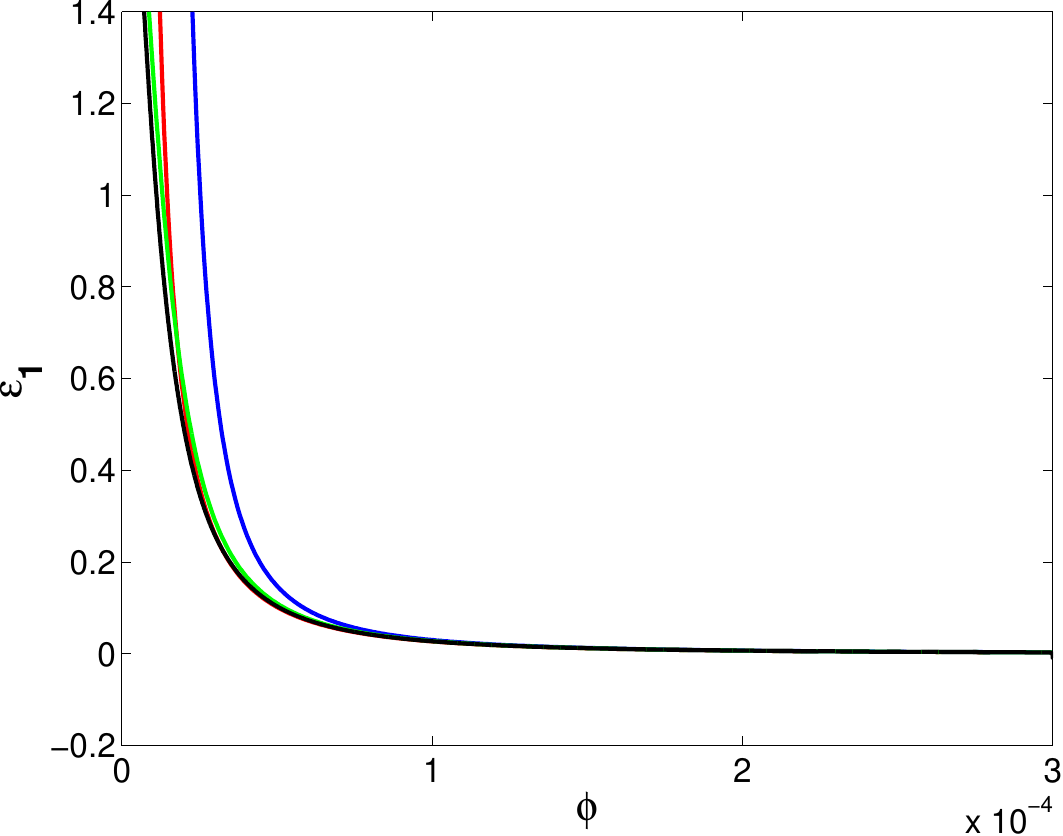}\quad
\includegraphics[scale=0.41]{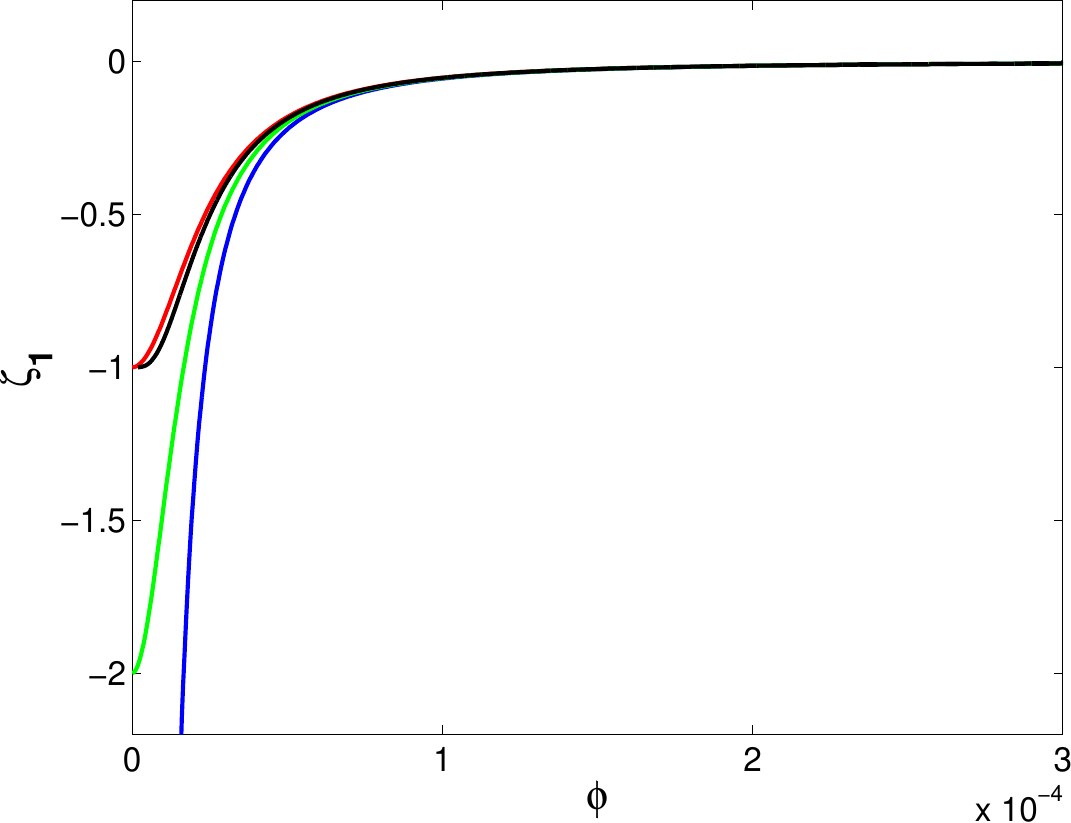}
\caption{The functions $\varepsilon_1(\phi)$ (left) and $\zeta_1(\phi)$ (right) for the model with the potential $V(\phi)=2\cdot 10^{8}\phi^4$ and the coupling function $F(\phi)=1+2.4\cdot 10^{9}\phi^2$. The black lines are the result of the numerical integration of the system (\ref{PhiSYSN}) with the initial data $\phi(N_0)=0.0003$, $\chi(N_0)=0$ at $N_0=-103$, the blue curves are obtained in the known approximation, the red curves are obtained in the approximation I, and the green curves are obtained in the approximation III.}
\label{Fig7}
\end{figure}

\begin{figure}
\includegraphics[scale=0.41]{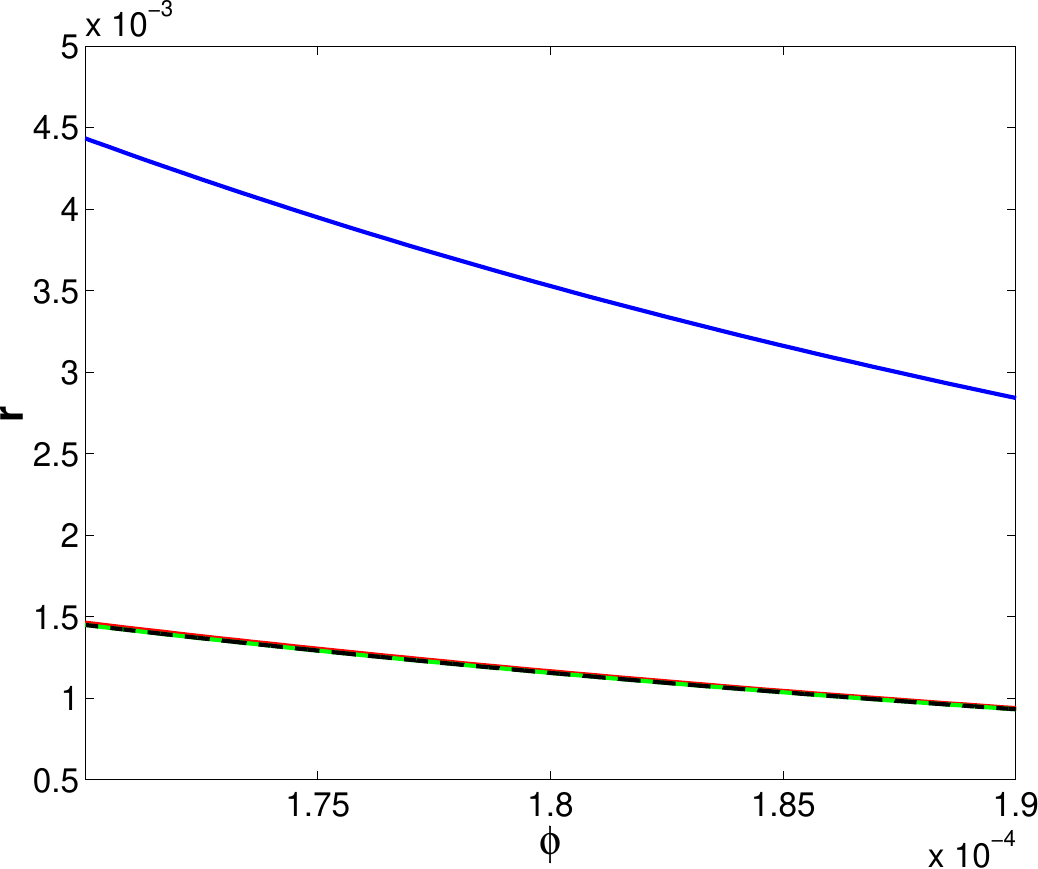}\quad
\includegraphics[scale=0.41]{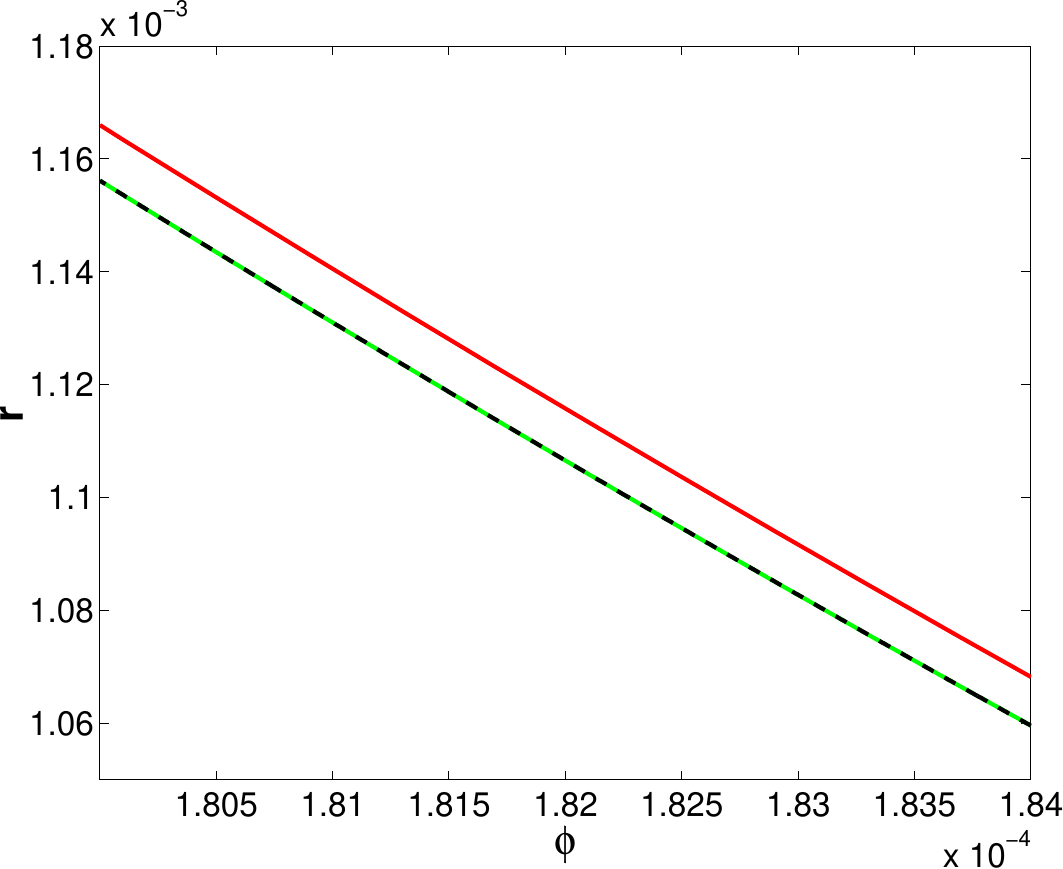}
\caption{The function $r(\phi)$ for the model with the potential $V(\phi)=2\cdot 10^{8}\phi^4$ and the coupling function $F(\phi)=1+2.4\cdot 10^{9}\phi^2$. The black lines are the result of the numerical integration of the system (\ref{PhiSYSN}), the blue curves are obtained in the known approximation, the red curves are obtained in the approximation I, and the green curves are obtained in the approximation~III.}
\label{Fig8}
\end{figure}

\begin{figure}
\includegraphics[scale=0.27]{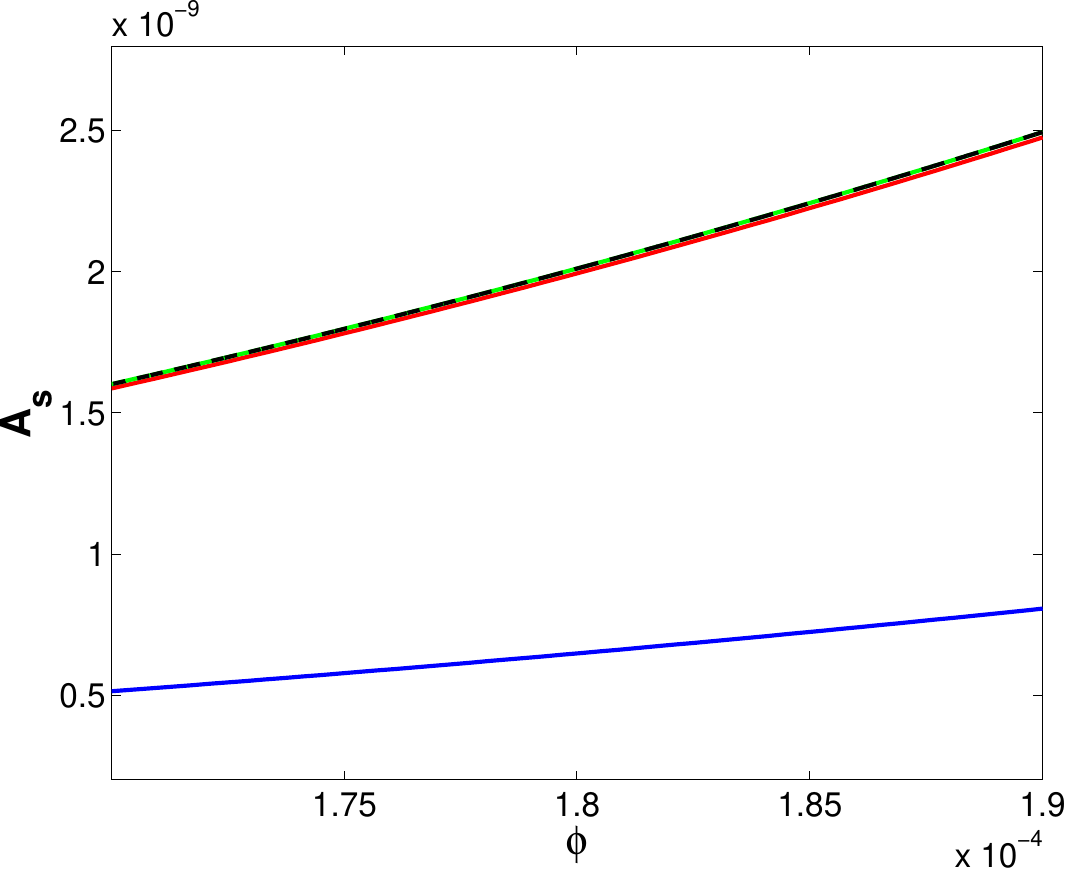}
\includegraphics[scale=0.27]{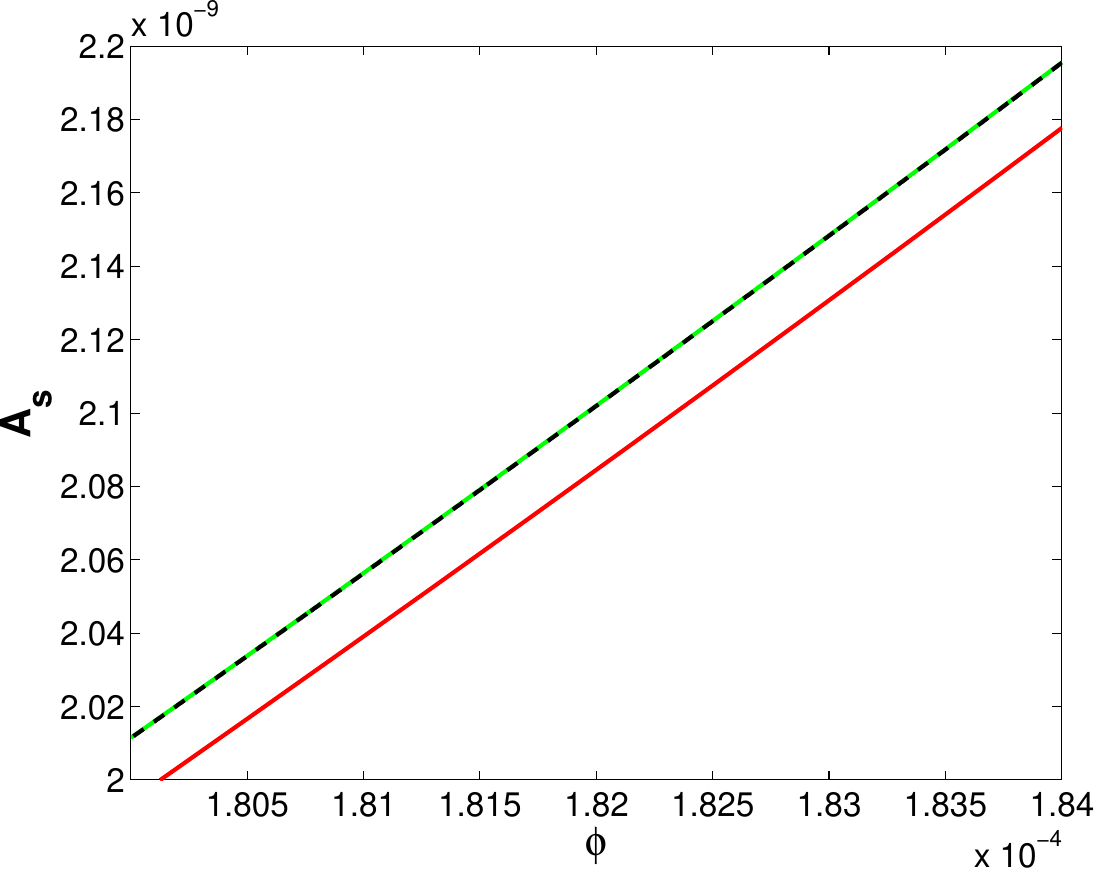}
\includegraphics[scale=0.27]{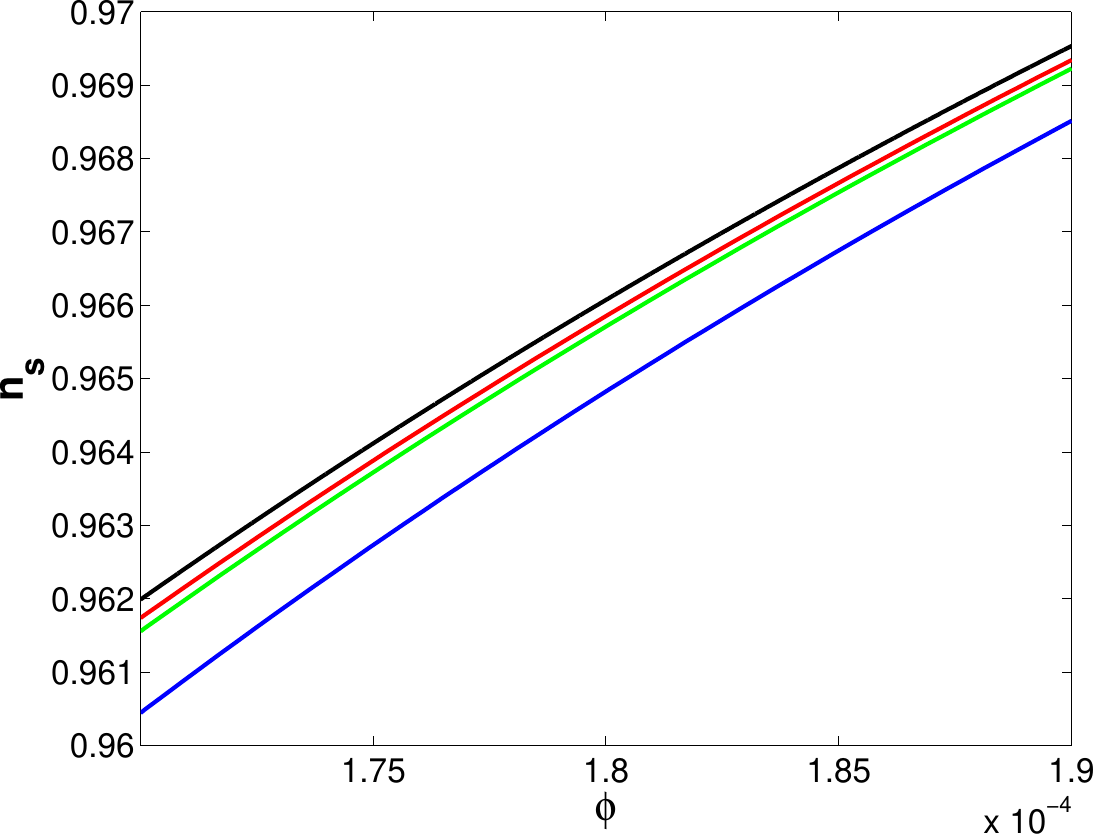}
\caption{The functions $A_s(\phi)$ (the left and center pictures) and $n_s(\phi)$ (the right picture) for the model with the potential $V(\phi)=2\cdot 10^{8}\phi^4$ and the coupling function $F(\phi)=1+2.4\cdot 10^{9}\phi^2$. The black lines are the result of the numerical integration of the system (\ref{PhiSYSN}), the blue curves are obtained in the known approximation, the red curves are obtained in the approximation I, and the green curves are obtained in the approximation III.}
\label{Fig9}
\end{figure}

    Similar results have been obtained also for large values of parameters, namely, $\xi=2.4\cdot 10^{9}$ and $\lambda=8\cdot 10^{8}$ (see Figs.~\ref{Fig7}--\ref{Fig9}). So, we conclude the proposed approximations I and III are essentially more accurate than the best known approximation.

\section{Conclusions and discussions}
~~~~The accuracy of the observation constraints on the inflationary parameters is increasing. It requires the use of more precise slow-roll approximations to construct inflationary models. In this paper, we propose new slow-roll approximations for models with a single scalar field nonminimally coupled to gravity.

    The main difference between the proposed and known approximations lies in the simplified form of Eq.~(\ref{equ00}). The new approximations are more accurate because they neglect only one term in this equation, while two summands were considered negligibly small in both standard approximations.

    The term $(1+\zeta_1)$ in the denominator can make the approximation II invalid, if the value of $\zeta_1$ comes close to $-1$ or even crosses this value. This situation does occur for the Higgs inflationary model. As one can see in Fig.~\ref{FigHiggs1}, an attempt to use the approximation II for the considered models with nonminimal coupling leads to $\zeta_1$ crossing $-1$, and then to inappropriate behavior of other variables. Therefore, we disregard this approximation for the models studied here. At the same time, the proposed approximations I and III are essentially more precise than the known slow-roll approximation. The proposed versions of the slow-roll approximation are not only more accurate at the end of inflation, but also give essentially more precise values of the tensor-to-scalar ratio~$r$ and the amplitude of scalar perturbations~$A_s$.

    We would like to point out that the new approximation II proposed in this paper can be considered as an analog of the corresponding approximation in the Einstein-Gauss-Bonnet gravity, which were proposed in our recent paper~\cite{Pozdeeva:2024ihc}. We remind a reader that for specific Gauss-Bonnet models studied in Ref.~\cite{Pozdeeva:2024ihc} the  approximation II works perfectly. However, as it was already pointed out in Ref.~\cite{Pozdeeva:2024ihc}, a term in the denominator containing a slow-roll parameter can make the  approximation II invalid. Such a situation did not happen in our previous paper~\cite{Pozdeeva:2024ihc}, but it is the case for the models analyzed in the present paper.

    Note, however, that we have described two ways of derivation of the approximation II. The way presented in Appendix B is similar to one proposed in the Gauss-Bonnet models, while the other one uses essentially the function $Y$, which corresponds to the Hubble parameter in the Einstein frame. The approximation II includes, in fact, one additional simplification step which can be lifted, resulting in the approximation III. In this new approximation, the factor $(1+\zeta_1)$ in the denominator is replaced by $(1+\zeta_1/2)$ and it has important consequences, since $\zeta_1$ cannot cross $-2$ in any case, and conditions that make the approximation II invalid cannot invalidate the approximation III. We have confirmed numerically that the approximation III usually work very well. However, for small values of the coupling constant $\xi$, the approximation I appears to be slightly more accurate than the approximation~III.

    As we have remarked, the approximation III is in fact the standard slow-roll approximation written in the Einstein frame and transformed back to the Jordan frame. Since there is no Einstein frame for the Einstein-Gauss-Bonnet gravity, the approximation III has no analog in such cosmological models. However, the fact that we arrived at this approximation without relying on the Einstein frame allows researchers to use it in more complicated models where nonminimal coupling is present, even if the Einstein frame does not exist for the entire theory due to additional terms, for example, the Gauss-Bonnet term. Another possible way of generalization of the considered model is to add the $F(R)$ term and to consider models that correspond to two field models in the Einstein frame. Such models are being actively investigated~\cite{Kaneda:2015jma,Ema:2017rqn,He:2018gyf,Gorbunov:2018llf,Canko:2019mud,Gundhi:2020zvb,Cecchini:2024xoq}.
        We plan to consider these models in future investigations.

\subsection*{Acknowledgements}

The study was conducted under the state assignment of Lomonosov Moscow State University.

\appendix

\section{The slow-roll parameters $\varepsilon_2(\phi)$ and $\zeta_2(\phi)$ in different approximations}

\subsection{The simplest slow-roll approximation}
~~~~In the simplest approximation, we get
\begin{equation}
\label{eps2apprS}
\begin{split}
\varepsilon_2(\phi)&= \frac{2F_{,\phi}V-FV_{,\phi}}{V}\\
&\times\left(\frac{F_{,\phi}V_{,\phi\phi}-F_{,\phi\phi}V}{FV_{,\phi}-F_{,\phi}V}+\frac{FV_{,\phi\phi}-F_{,\phi}V_{,\phi}-2F_{,\phi\phi}V}{FV_{,\phi}-2F_{,\phi}V}-\frac{F_{,\phi}}{F}-2\frac{V_{,\phi}}{V}\right)
\end{split}
\end{equation}
and
\begin{equation}
\label{zeta2apprS}
\zeta_2(\phi)= \frac{2F_{,\phi}V-FV_{,\phi}}{V}\left(\frac{F_{,\phi\phi}}{F_{,\phi}}+\frac{F_{,\phi}V_{,\phi}+2F_{,\phi\phi}V-FV_{,\phi\phi}}{2F_{,\phi}V-FV_{,\phi}}-\frac{F_{,\phi}}{F}-\frac{V_{,\phi}}{V}\right).
\end{equation}

\subsection{The known more accurate slow-roll approximation}
~~~~In this approximation,
\begin{equation}
\label{eps2appr1a}
\begin{split}
\varepsilon_2(\phi)&= \frac{2F(2F_{,\phi}V-FV_{,\phi})}{V(2F+3F_{,\phi}^2)}\\
&\times\left(\frac{F_{,\phi}V_{,\phi\phi}-F_{,\phi\phi}V}{FV_{,\phi}-F_{,\phi}V}+\frac{FV_{,\phi\phi}-F_{,\phi}V_{,\phi}-2F_{,\phi\phi}V}{FV_{,\phi}-2F_{,\phi}V}-2\frac{V_{,\phi}}{V}-\frac{2F_{,\phi}(1+3F_{,\phi\phi})}{2F+3F_{,\phi}^2}\right)
\end{split}
\end{equation}

and

\begin{equation}
\label{zeta2appr1a}
\begin{split}
\zeta_2(\phi)&= \frac{2F(2F_{,\phi}V-FV_{,\phi})}{V(2F+3F_{,\phi}^2)}\\
&\times\left(\frac{F_{,\phi\phi}}{F_{,\phi}}+\frac{V_{,\phi}F_{,\phi}+2VF_{,\phi\phi}-FV_{,\phi\phi}}{2VF_{,\phi}-FV_{,\phi}}-\frac{V_{,\phi}}{V}-\frac{2F_{,\phi}(1+3F_{,\phi\phi})}{2F+3F_{,\phi}^2}\right).
\end{split}
\end{equation}

\subsection{New slow-roll approximation I}
~~~~In this approximation,
\begin{equation}
\begin{array}{l}
\label{eps2appr1}
\varepsilon_2(\phi)= \frac{2F(2F_{,\phi}V-FV_{,\phi})}{2FV+2FF_{,\phi}V_{,\phi}-F_{,\phi}^2V}\\
\\\times\left\lbrace \frac{F_{,\phi}}{F}+\frac{2F_{,\phi\phi}V+F_{,\phi}V_{,\phi}-FV_{,\phi\phi}}{2F_{,\phi}V-V_{,\phi}F}-\frac{2FV_{,\phi}+2F_{,\phi}V+F_{,\phi}^2V_{,\phi}
+2FF_{,\phi\phi}V_{,\phi}+2FF_{,\phi}V_{,\phi\phi}-2F_{,\phi}F_{,\phi\phi}V}{2FV+2FF_{,\phi}V_{,\phi}-F_{,\phi}^2V}\right. \\ \\
 \left. +{\left[\frac{2FV_{,\phi}+2F_{,\phi}V+F_{,\phi}^2V_{,\phi}+2FF_{,\phi\phi}V_{,\phi}+2FF_{,\phi}V_{,\phi\phi}-2F_{,\phi}F_{,\phi\phi}V}{2FV+2FF_{,\phi}V_{,\phi}-F_{,\phi}^2V}-\frac{F_{,\phi}}{F}
 -\frac{2F_{,\phi}(1+3F_{,\phi\phi})}{2F+3F_{,\phi}^2}\right]}^{-1}\right. \\ \\\left. \times\left(\frac{2FV_{,\phi\phi}+4F_{,\phi}V_{,\phi}+2F_{,\phi\phi}V+3F_{,\phi}^2V_{,\phi\phi}+2F_{,\phi}F_{,\phi\phi}V_{,\phi}-2F_{,\phi\phi}^2V-2F_{,\phi}F_{,\phi\phi\phi}V+2FF_{,\phi\phi\phi}V_{,\phi}+4FF_{,\phi\phi}V_{,\phi\phi}+2FF_{,\phi}V_{,\phi\phi\phi}}{2FV+2FF_{,\phi}V_{,\phi}-F_{,\phi}^2V}\right.\right. \\ \\ \left.\left. -\frac{F_{,\phi\phi}}{F}+\frac{F_{,\phi}^2}{F^2}-\frac{{(2FV_{,\phi}+2F_{,\phi}V-F_{,\phi}^2V_{,\phi}-2F_{,\phi}F_{,\phi\phi}V+2F_{,\phi}^2V_{,\phi}+2FF_{,\phi\phi}V_{,\phi}+2FF_{,\phi}V_{,\phi\phi})}^2}{{(2FV+2FF_{,\phi}V_{,\phi}-F_{,\phi}^2V)}^2}\right.\right. \\ \\ \left.\left. -\frac{2(F_{,\phi\phi}+3F_{,\phi\phi}^2+3F_{,\phi}F_{,\phi\phi\phi})}{2F+3F_{,\phi}^2}+\frac{4F_{,\phi}^2{(1+3F_{,\phi\phi})}^2}{{(2F+3F_{,\phi}^2)}^2}\right)\right\rbrace,
\end{array}
\end{equation}

\begin{equation}
\label{zeta2appr1}
\begin{split}
\zeta_2(\phi)&=\frac{d\zeta_1}{\zeta_1dN}=\frac{2F(2F_{,\phi}V-FV_{,\phi})}{2FV+2FF_{,\phi}V_{,\phi}-F_{,\phi}^2V}\\
&\times\left[\frac{F_{,\phi\phi}}{F_{,\phi}}+\frac{2F_{,\phi\phi}V+F_{,\phi}V_{,\phi}-FV_{,\phi\phi}}{2F_{,\phi}V-FV_{,\phi}}\right.\\
&\left.{}-\frac{2FV_{,\phi}+2F_{,\phi}V+F_{,\phi}^2V_{,\phi}+2FF_{,\phi\phi}V_{,\phi}+2FF_{,\phi}V_{,\phi\phi}-2F_{,\phi}F_{,\phi\phi}V}{2FV+2FF_{,\phi}V_{,\phi}-F_{,\phi}^2V}\right].
\end{split}
\end{equation}

\subsection{New slow-roll approximation II}
~~~~In this approximation,
\begin{equation}
\begin{array}{l}
\label{eps2appr2}
\varepsilon_2(\phi)= \frac{2F(2F_{,\phi}V-FV_{,\phi})}{2FV+FF_{,\phi}V_{,\phi}+F_{,\phi}^2V}\\
\\\times\left\lbrace  \frac{F_{,\phi}}{F}+\frac{2F_{,\phi\phi}V+F_{,\phi}V_{,\phi}-FV_{,\phi\phi}}{2F_{,\phi}V-FV_{,\phi}}-\frac{2FV_{,\phi}+2F_{,\phi}V+FF_{,\phi}V_{,\phi\phi}+2F_{,\phi}^2V_{,\phi}+FF_{,\phi\phi}V_{,\phi}+2F_{,\phi}F_{,\phi\phi}V}{2FV
+FF_{,\phi}V_{,\phi}+F_{,\phi}^2V}\right. \\
\\\left. +\left[\frac{V_{,\phi}}{V}+\frac{2FV_{,\phi}+2F_{,\phi}V+FF_{,\phi}V_{,\phi\phi}+2F_{,\phi}^2V_{,\phi}+FF_{,\phi\phi}V_{,\phi}+2F_{,\phi}F_{,\phi\phi}V}{2FV+FF_{,\phi}V_{,\phi}+F_{,\phi}^2V}\right.\right. \\
\\ \left.\left. -\frac{F_{,\phi}}{F}-\frac{2FV_{,\phi}+2F_{,\phi}V-FF_{,\phi}V_{,\phi\phi}+4F_{,\phi}^2V_{,\phi}-FF_{,\phi\phi}V_{,\phi}+10F_{,\phi}F_{,\phi\phi}V}{2FV-FF_{,\phi}V_{,\phi}+5F_{,\phi}^2V}\right]^{-1}\right. \\
\\\left.\times\left(\frac{2FV_{,\phi\phi}+4F_{,\phi}V_{,\phi}+2F_{,\phi\phi}V+FF_{,\phi}V_{,\phi\phi\phi}+3F_{,\phi}^2V_{,\phi\phi}+2FF_{,\phi\phi}V_{,\phi\phi}+7F_{,\phi}F_{,\phi\phi}V_{,\phi}+FF_{,\phi\phi\phi}V_{,\phi}
+2F_{,\phi\phi}^2V+2F_{,\phi}F_{,\phi\phi\phi}V}{2FV+FF_{,\phi}V_{,\phi}+F_{,\phi}^2V}\right.\right. \\
\\\left.\left. {}-\frac{{(2FV_{,\phi}+2F_{,\phi}V+FF_{,\phi}V_{,\phi\phi}+2F_{,\phi}^2V_{,\phi}+FF_{,\phi\phi}V_{,\phi}+2F_{,\phi}F_{,\phi\phi}V)}^2}{{(2FV+FF_{,\phi}V_{,\phi}
+F_{,\phi}^2V)}^2}+\frac{V_{,\phi\phi}}{V}-\frac{{V_{,\phi}}^2}{V^2}-\frac{F_{,\phi\phi}}{F}+\frac{F_{,\phi}^2}{F^2}\right.\right. \\
\\ \left.\left. {} - \frac{2FV_{,\phi\phi}+4F_{,\phi}V_{,\phi}+2F_{,\phi\phi}V-FF_{,\phi}V_{,\phi\phi\phi}+3F_{,\phi}^2V_{,\phi\phi}-2FF_{,\phi\phi}
V_{,\phi\phi}+17F_{,\phi}F_{,\phi\phi}V_{,\phi}-FF_{,\phi\phi\phi}V_{,\phi}+10F_{,\phi\phi\phi}F_{,\phi}V+10F_{,\phi\phi}^2V}{2FV-FF_{,\phi}V_{,\phi}+5F_{,\phi}^2V}\right.\right. \\
\\ \left.\left. {} -\frac{{\left(2FV_{,\phi}+2F_{,\phi}V-FF_{,\phi}V_{,\phi\phi}+4F_{,\phi}^2V_{,\phi}-FF_{,\phi\phi}V_{,\phi}+10F_{,\phi}F_{,\phi\phi}V\right)}^2}{{(2FV-FF_{,\phi}V_{,\phi}+5F_{,\phi}^2V)}^2}\right)\right
\rbrace\,.
\end{array}
\end{equation}

\begin{equation}
\begin{split}
\label{zeta2appr3}
\zeta_2(\phi)&= \frac{2F(2F_{,\phi}V-FV_{,\phi})}{2FV+FF_{,\phi}V_{,\phi}+F_{,\phi}^2V} \\ &\times\left[\frac{F_{,\phi\phi}}{F_{,\phi}}+\frac{F_{,\phi}V_{,\phi}+2F_{,\phi\phi}V-FV_{,\phi\phi}}{2F_{,\phi}V-FV_{,\phi}}\right.
\\ &\left.{}-\frac{2FV_{,\phi}+2F_{,\phi}V+FF_{,\phi}V_{,\phi\phi}+2F_{,\phi}^2V_{,\phi}+FF_{,\phi\phi}V_{,\phi}+2F_{,\phi}F_{,\phi\phi}V}{2FV+FF_{,\phi}V_{,\phi}+F_{,\phi}^2V}\right].
\end{split}
\end{equation}

\subsection{New slow-roll approximation III}
~~~~In this approximation,
\begin{equation}
\begin{array}{l}
\label{eps2appr3}
\varepsilon_2(\phi)= \frac{2F(2F_{,\phi}V-FV_{,\phi})}{2FV+FF_{,\phi}V_{,\phi}+F_{,\phi}^2V}\\
\\\times\left\lbrace \frac{F_{,\phi}}{F}+\frac{2F_{,\phi\phi}V+F_{,\phi}V_{,\phi}-FV_{,\phi\phi}}{2F_{,\phi}V-FV_{,\phi}}-\frac{2FV_{,\phi}+2F_{,\phi}V+FF_{,\phi}V_{,\phi\phi}+2F_{,\phi}^2V_{,\phi}+FF_{,\phi\phi}V_{,\phi}
+2F_{,\phi}F_{,\phi\phi}V}{2FV+FF_{,\phi}V_{,\phi}+F_{,\phi}^2V}\right. \\
\\ \left. +\left[-\frac{V_{,\phi}}{V}+\frac{2\left(2FV_{,\phi}+2F_{,\phi}V+FF_{,\phi}V_{,\phi\phi}+2F_{,\phi}^2V_{,\phi}+FF_{,\phi\phi}V_{,\phi}+2F_{,\phi}F_{,\phi\phi}V\right)}{2FV+FF_{,\phi}V_{,\phi}+F_{,\phi}^2V}
-\frac{F_{,\phi}}{F}-\frac{{4F_{,\phi}\left(1+3F_{,\phi\phi}\right)}}{2F+3F_{,\phi}^2}\right]^{-1}\right. \\
\\\left.\times\left(\frac{2\left(2FV_{,\phi\phi}+4F_{,\phi}V_{,\phi}+2F_{,\phi\phi}V+FF_{,\phi}V_{,\phi\phi\phi}+3F_{,\phi}^2V_{,\phi\phi}+2FF_{,\phi\phi}V_{,\phi\phi}+7F_{,\phi}F_{,\phi\phi}V_{,\phi}
+FF_{,\phi\phi\phi}V_{,\phi}+2F_{,\phi\phi}^2V+2F_{,\phi}F_{,\phi\phi\phi}V\right)}{2FV+FF_{,\phi}V_{,\phi}+F_{,\phi}^2V}\right.\right. \\
\\ \left.\left. -\frac{{2\left(2FV_{,\phi}+2F_{,\phi}V+FF_{,\phi}V_{,\phi\phi}+2F_{,\phi}^2V_{,\phi}+FF_{,\phi\phi}V_{,\phi}+2F_{,\phi}F_{,\phi\phi}V\right)}^2}{{(2FV+FF_{,\phi}V_{,\phi}
+F_{,\phi}^2V)}^2}-\frac{V_{,\phi\phi}}{V}+\frac{{V_{,\phi}}^2}{V^2}-\frac{F_{,\phi\phi}}{F}+\frac{F_{,\phi}^2}{F^2}\right.\right. \\
\\ \left.\left. -\frac{4(F_{,\phi\phi}+3F_{,\phi\phi}^2+3F_{,\phi}F_{,\phi\phi\phi})}{2F+3F_{,\phi}^2}+\frac{8F_{,\phi}^2{(1+3F_{,\phi\phi})}^2}{{(2F+3F_{,\phi}^2)}^2}\right)\right\rbrace \,.
\end{array}
\end{equation}

The value of $\zeta_2(\phi)$ is given by Eq.~(\ref{zeta2appr3}).

\section{Another way to get approximation II}
We neglect the term proportional to $\zeta_1^2$ in Eq.~(\ref{equ00eps1zeta1zeta2}) and obtain
\begin{equation}
\label{00appr2}
H^2\approx\frac{V}{3F(1+\zeta_1)}.
\end{equation}

    Also, we get
\begin{equation}
\label{dotHappr2}
\dot H=\frac{\dot\phi}{2H}\frac{d(H^2)}{d\phi}\approx\frac{\zeta_1 V}{6F_{,\phi}(1+\zeta_1)}\left(\frac{V_{,\phi}}{V}-\frac{F_{,\phi}}{F}-\frac{\zeta_2F_{,\phi}}{F(1+\zeta_1)}\right).
\end{equation}

    Neglecting the term proportional to $\zeta_1\zeta_2$, we find
\begin{equation}
\label{dotHappr2a}
\dot{H}\approx\frac{\zeta_1 V}{6F_{,\phi}(1+\zeta_1)}\left(\frac{V_{,\phi}}{V}-\frac{F_{,\phi}}{F}\right)=
\frac{\zeta_1}{2}H^2\left(\frac{V_{,\phi}}{V}\frac{F}{F_{,\phi}}-1\right).
\end{equation}

    Substituting (\ref{00appr2}) and (\ref{dotHappr2a}) to (\ref{GBappr1}), we find a linear equation in $\zeta_1$ with the following solution:
\begin{equation}
\label{zeta1appr2}
\zeta_1(\phi)=\frac{2F_{,\phi}(2F_{,\phi}V-FV_{,\phi})}{2FV+FF_{,\phi}V_{,\phi}+F_{,\phi}^2V}.
\end{equation}

    Now, we apply (\ref{zeta1appr2}) and find
\begin{equation}
\chi(\phi)\approx\frac{2F\left(2F_{,\phi}V-FV_{,\phi}\right)}{2FV+FF_{,\phi}V_{,\phi}+F_{,\phi}^2V},
\end{equation}
and
\begin{equation}
H^2(\phi)\approx\frac{V}{3F(1+\zeta_1)}=\frac{V\left(2FV+FF_{,\phi}V_{,\phi}+F_{,\phi}^2V\right)}{3F\left(2FV-FF_{,\phi}V_{,\phi}+5F_{,\phi}^2V\right)}.
\end{equation}

So, we come to the approximation II.

\bibliographystyle{JHEP}
\bibliography{NonminScalarINSPIRE}

\end{document}